\documentclass[12pt]{article}
\usepackage{bbm}
\usepackage{color}
\usepackage{authblk}
\usepackage{latexsym}
\usepackage{epsfig,amssymb,euscript, mathrsfs,cite}
\usepackage{amsmath}
\usepackage{slashed}
\usepackage{cancel}
\usepackage{verbatim}
\usepackage{mathrsfs} 
\usepackage[normalem]{ulem}
\usepackage{enumerate}
\definecolor{MyDarkBlue}{rgb}{0.15,0.15,0.45}
\usepackage[linktocpage=true]{hyperref}
\hypersetup{
colorlinks=true,
citecolor=MyDarkBlue,
linkcolor=MyDarkBlue,
urlcolor=MyDarkBlue,
pdfauthor={},
pdftitle={},
pdfsubject={hep-th}
}
\def\v{\vec{v}}
\def\w{\vec{w}}

\newsavebox{\ns}
\newsavebox{\dbrane}
\newsavebox{\dbshort}

\def\be{\begin{equation}}
\def\ee{\end{equation}}
\def\bea{\begin{eqnarray}}
\def\eea{\end{eqnarray}}

\newcommand{\nn}{\nonumber\\}

\newcommand{\ii}{\mathrm{i}}

\newcommand{\R}{\mathbb{R}}
\newcommand{\Z}{\mathbb{Z}}

\newcommand\diff{\mathrm{d}}

\newcommand{\ex}{\mathrm{e}}
\newcommand{\vol}{\mathrm{vol}}
\newcommand{\Vol}{\mathrm{Vol}}

\newcommand{\cZ}{\mathscr{Z}}
\newcommand{\Ssusy}{S_{\mathrm{SUSY}}}

\newlength{\sswidth}

\numberwithin{equation}{section}       

\begin{document}

\begin{titlepage}



\vskip 1cm
\vskip 1cm

\begin{center}


{\Large \bf Sasaki-Einstein Geometry, GK Geometry\\[3mm] 
and the AdS/CFT correspondence}

\vskip 1cm
Jerome P. Gauntlett$^{\mathrm{a}}$, Dario Martelli$^{\mathrm{b,c}}$ and James Sparks$^{\mathrm{d}}$

\vskip 1cm

\vskip 0.2cm

${}^{\mathrm{a}}$\textit{Blackett Laboratory, Imperial College, \\
Prince Consort Rd., London, SW7 2AZ, U.K.\\}

\vskip 0.2cm
${}^{\mathrm{b}}$\textit{Dipartimento di Matematica
 ``Giuseppe Peano'', \\
 Universit\`a di Torino, Via Carlo Alberto 10, 10123 Torino, Italy\\}

\vskip 0.2cm

${}^{\mathrm{c}}$\textit{INFN, Sezione di Torino, Via Pietro Giuria 1, 10125 Torino, Italy \\}
 
\vskip 0.2cm

${}^{\mathrm{d}}$\textit{Mathematical Institute, University of Oxford,\\
Andrew Wiles Building, Radcliffe Observatory Quarter,\\
Woodstock Road, Oxford, OX2 6GG, U.K.\\}

\vskip 0.2 cm

\end{center}

\vskip 0.9 cm

\begin{abstract}
\noindent  
We review various aspects of Sasaki-Einstein and GK geometry, emphasising their similarities, interconnections and significance for the AdS/CFT correspondence.
In particular, we highlight the key role that physical considerations have played in formulating geometric 
extremization principles, which have been instrumental in both understanding the geometry and identifying 
the corresponding dual field theories.

\vskip 4cm
{\it \noindent Invited contribution to the book ``Half a Century of Supergravity'', editors A.~Ceresole and G.~Dall'Agata.}

\end{abstract}
\end{titlepage}

\pagestyle{plain}
\setcounter{page}{1}
\newcounter{bean}
\baselineskip18pt



\section{Historical overview}

The AdS/CFT correspondence, discovered by Juan Maldacena in 1997 \cite{Maldacena:1997re}, is one of the most influential discoveries within string/M-theory. It has revealed profound interconnections between quantum field theory and quantum gravity as well as provided striking connections with geometry.
Two of the original examples are the $AdS_5\times S^5$ solution 
of type IIB supergravity and the $AdS_4\times S^7$ solution of $D=11$ supergravity, both of which preserve maximal
supersymmetry. A key observation is that these solutions arise as the near horizon geometry of
stacks of D3-branes and M2-branes in flat spacetime, respectively. In the former case this led to the
identification of the dual SCFT as $d=4$ $\mathcal{N}=4$ super-Yang-Mills (SYM) theory, while in the latter case it took another ten years
before the connection with $d=3$ ABJM theory was made \cite{Aharony:2008ug}. 

Shortly after the appearance of Maldacena's paper it was realised that additional supersymmetric 
examples of the correspondence can be realised by placing D3-branes or M2-branes at the apex of Calabi-Yau (CY) cones \cite{Klebanov:1998hh,Acharya:1998db,Morrison:1998cs}. 
The cross section of a CY cone 
is, by definition, a Sasaki-Einstein (SE) geometry
and the brane construction directly leads to supersymmetric $AdS_5\times SE_5$ solutions of type IIB supergravity, 
generically dual to ${\cal N}=1$ SCFTs in $d=4$, and supersymmetric $AdS_4\times SE_7$ solutions of $D=11$ supergravity, generically dual to ${\cal N}=2$ SCFTs in $d=3$. 

Until 2004, however, only a few SE metrics were explicitly known. 
These examples, all of which are homogeneous spaces, had been identified in the physics literature in the context of 
Freund-Rubin type vacua \cite{Freund:1980xh}. A few homogeneous examples exist in $D=7$
(e.g. \cite{Castellani:1983yg}).
However, since the connection between the $AdS_4\times S^7$ solution and ABJM theory was not yet established in 1998, 
it was unclear how to gain further insight into the dual $d=3$ SCFTs. 
In $D=5$, apart from the round $S^5$, the only other explicit solution was 
$T^{1,1}$ \cite{Romans:1984an}, 
which is topologically $S^2\times S^3$ with a homogeneous
 SE metric. This example was particularly significant since the dual ${\cal N}=1$ SCFT was identified as a specific quiver gauge theory \cite{Klebanov:1998hh}. 

The discovery of the infinite class of explicit $Y^{p,q}$ metrics in 2004 \cite{Gauntlett:2004yd} had a significant impact
on AdS/CFT. The $Y^{p,q}$ are co-homogeneity one SE metrics on $S^2\times S^3$, labelled by two relatively prime, positive integers $p,q$. 
The discovery of $Y^{p,q}$ was also somewhat of a surprise to the math community; for example they provided
the very first examples of SE metrics in the irregular class\footnote{The \emph{irregular} class is when the
Killing vector dual to the R-symmetry has orbits which are generically $\mathbb{R}$.}
 which had been conjectured not to exist at all \cite{CheegerJeff1994Otcs}.
Soon afterwards, the SCFTs dual to $Y^{p,q}$ were identified 
as quiver gauge theories \cite{Benvenuti:2004dy}. A key 
ingredient in this identification was the newly discovered 
principle of $a$-maximization for $d=4$ $\mathcal{N}=1$ SCFTs, which allows one to identify the 
R-symmetry of the SCFT and hence its central charge, by an extremization principle based on anomaly considerations  \cite{Intriligator:2003jj}.
These discoveries catalysed a number of further developments.

One fruitful theme has been to study the 
implications of general features of the dual SCFTs when translated into geometric terms. In turn the geometric understanding gives rise to new insights into the strongly coupled SCFTs. 
For example, since the $a$ central charge of the $d=4$ SCFT is inversely proportional to the volume of the corresponding 
$SE_5$,
$a$-maximization in field theory suggested that there could be a geometric extremization principle for obtaining the volume of $SE_5$ metrics. Indeed, there is such a notion of volume extremization 
\cite{Martelli:2005tp,Martelli:2006yb}.
In fact this 
volume 
extremization principle applies to SE metrics in any dimension. For the particular
case of $SE_7$ this result 
indicated that there is some kind of extremization principle at work for the dual $d=3$ SCFTs. 
This was later 
shown to be associated with the extremization of the free energy of the $d=3$ SCFT when placed on $S^3$ \cite{Jafferis:2010un}, thus completing the circle.
 
In hindsight, there are several ways in which the $Y^{p,q}$ metrics could have been discovered, including 
directly constructing them as SE metrics using a co-homogeneity one ansatz and then solving a resulting set of ODEs. 
Interestingly, however, they were 
discovered indirectly as a by-product of a general programme of classifying supersymmetric supergravity solutions. The basic idea of the ``G-structure programme", which was founded at Queen Mary, University of London with a series of papers in the early 2000's, 
is to systematically extract the geometric information
encoded in the Killing spinor equations by studying the differential forms 
that can be constructed as spinor bilinears.
These forms satisfy algebraic conditions, arising from Fierz identities, as well as differential conditions arising from the 
Killing spinor equations. More geometrically, the Killing spinor, or equivalently the differential forms, defines a G-structure and the differential conditions constrain the intrinsic torsion.
These techniques have been used in various contexts: early G-structure papers \cite{Gauntlett:2001ur,Gauntlett:2002sc,Gauntlett:2003cy} were motivated by 
wrapped brane geometries and flux compactifications; 
the classification of the most general supersymmetric solutions of $D=11$ supergravity was carried out in
\cite{Gauntlett:2002fz,Gauntlett:2003wb}; and
the classification of general supersymmetric solutions of minimal $D=5$ supergravity, both ungauged and gauged, in \cite{Gauntlett:2002nw,Gauntlett:2003fk}
which impacted both on studies of black holes in string/M-theory and the fuzzball programme e.g. \cite{Elvang:2004rt, Gauntlett:2004qy, Bena:2007kg}.

These G-structure techniques were also used to initiate the programme of systematically characterizing the geometry 
associated with supersymmetric AdS/CFT \cite{Gauntlett:2004zh}.
The goal is to study the most general class of SCFTs in any spacetime dimension $d\le 6$, 
which have a supergravity dual. This translates into studying the most general $AdS_{d+1}\times M$ solutions of 
$D=10,11$
supergravity, with 
metric and fluxes preserving the isometries of $AdS_{d+1}$. 
One assumes that they admit
a Killing spinor, then utilises the G-structure technology to precisely characterize the geometry on $M$. 
In general one needs to develop, in a case by case manner, new tools to study the geometry arising on $M$. In some cases
it is 
possible to find explicit solutions which are invariably helpful in making further progress.

The focus of \cite{Gauntlett:2004zh} was the most general class of supersymmetric $AdS_5\times M_6$ solutions of $D=11$ supergravity and hence the most general class of $d=4$ SCFTs with a $D=11$ $AdS_5$ dual. Having elucidated a precise characterization of $M_6$, infinite classes of explicit solutions were also found. 
A
sub-class of such solutions have a $T^2$ isometry;
this allows one to dimensionally reduce 
and T-dualise to obtain $AdS_5$ solutions of type IIB supergravity and these are precisely the $AdS_5\times Y^{p,q}$ solutions of \cite{Gauntlett:2004yd}.

The classification of supersymmetric AdS solutions, started in \cite{Gauntlett:2004zh}, has been extensively developed and continues to be an active area
of research. Many cases have now been analysed with some proving to be more tractable than others. 
The classification
of $AdS_3\times Y_7$ solutions with non-vanishing five-form flux was analysed in
\cite{Kim:2005ez} and these are dual to $\mathcal{N}=(0,2)$ SCFTs in $d=2$. In addition, $AdS_2\times Y_9$ solutions of $D=11$ with electric four-form flux were analysed in \cite{Kim:2006qu}; one way these arise is as the near horizon limit of supersymmetric black holes in $AdS_4$.
The geometry on $Y_7$ and $Y_9$ was further elucidated in \cite{Gauntlett:2007ts}. Moreover, it was also shown that it can be 
extended to arbitrary odd dimensions and is referred to as GK geometry.  

GK geometry has some striking similarities with SE geometry. Furthermore, SE geometry enters GK geometry in a physically important way and so
it is natural to discuss them together. Perhaps of most significance is that there is an extremization principle for GK geometry,
discovered in 2018 \cite{Couzens:2018wnk}, and, as in the SE case, 
leads to a deeper understanding of
GK geometry as well as the dual SCFTs e.g.\cite{Gauntlett:2018dpc,Hosseini:2019use,Hosseini:2019ddy,Gauntlett:2019roi,Kim:2019umc}. In particular, the principle allows one to compute quantities of physical interest without having explicit solutions, just inputting some topological information and assuming that the solutions exist.
The new extremization principle was
again 
inspired by field theory. The $c$-extremization principle \cite{Benini:2012cz,Benini:2013cda} for
$d=2$ $\mathcal{N}=(0,2)$ SCFTs suggested that there could be an extremization principle for $Y_7$, associated with
the $AdS_3\times Y_7$ solutions, but the principle found in \cite{Couzens:2018wnk} is applicable to GK geometry in any dimension.
In particular, it is associated with $AdS_2\times Y_9$ solutions, where there is a close association with
$\mathcal{I}$-extremization and the programme of obtaining a microstate interpretation for the entropy 
of supersymmetric and asymptotically $AdS_4$ black holes \cite{Benini:2015eyy}. Indeed a matching of black entropy computed using GK geometry
with the topologically twisted index has now been achieved for
infinite classes of examples in \cite{Hosseini:2019ddy,Gauntlett:2019roi}.

In the remainder of this article we review various aspects of SE and GK geometry.
A more detailed review of SE geometry can be found in \cite{Boyer:2008era,Sparks:2010sn}.

\section{Sasaki-Einstein  geometry}

\subsection{Sasaki-Einstein  geometry}
SE geometry is defined on an odd-dimensional manifold, $Y_{2n+1}$, with $n\ge 1$. The metric, being Einstein, satisfies
equations of motion that can be derived from the Einstein-Hilbert action 
\begin{align}\label{action1}
S 
 =  \int_{Y_{2n+1}} \left[R_{2n+1} -2n (2n-1)\right]\vol_{2n+1}~.
\end{align}
The geometry also admits certain Killing spinors $\epsilon$ satisfying
\begin{align}
\Big(\nabla_a-\frac{\ii}{2}\gamma_a\Big)\epsilon&=0\,.
\end{align}

The 
cases relevant for physics are $Y_5$ and $Y_7$, when $n=2$ and $n=3$. 
When $n=2$ we have
supersymmetric $AdS_5\times Y_5$ solutions of type IIB supergravity of the form 
\begin{align}\label{typeiibsolsse}
\diff s^2_{10} &= L^2 \left(\diff s^2({AdS_5}) + \diff s^2({Y_5})\right)~,\nn
F_5 &= -L^4\left(\vol_{{AdS}_5} + \vol(Y_5)\right)~,
\end{align}
where $\diff s^2({{AdS}_5})$ has unit radius. 
Flux quantization of the five-form fixes 
$L$
via 
$L^4=(2\pi)^4 g_s(\alpha')^2N/4\Vol(Y_5)$, with $N$ an integer. 
These solutions are dual to $d=4$ SCFTs with
$\mathcal{N}=1$ supersymmetry, which necessarily have an R-symmetry.  A key observable is the $a$ central charge of the
dual field theory, which in holography is given by 
\begin{align}\label{acentse}
a=\frac{\pi^3}{4\Vol(Y_5)}N^2\, .
\end{align}
Similarly, when $n=3$ we obtain supersymmetric $AdS_4\times Y_7$
solutions of $D=11$ supergravity
of the form
\begin{align}\label{ansatzd11se}
\diff s^2_{11} &= L^2 \Big(\frac{1}{4}\diff s^2({{AdS}_4}) + \diff s^2(Y_{7})\Big)~,\nn
G_4 &= \frac{3}{8}L^3\vol_{{AdS}_4} ~,
\end{align}
where $\diff s^2({{AdS}_4})$ has unit radius. 
Flux quantization
implies 
$L^6=(2\pi \ell_p)^6 N/6 \Vol(Y_7)$, with $N$ an integer.
These solutions are dual to $d=3$ SCFTs with $\mathcal{N}=2$
supersymmetry with an R-symmetry.  
A key observable is the free energy of the SCFT on $S^3$, given in holography by 
\begin{align}\label{fexps3}
F_{S^3}= \sqrt{\frac{2\pi^6}{27\Vol(Y_7)}}N^{3/2}\, .
\end{align}

The Killing spinor equations imply the metric on $Y_{2n+1}$ has a unit norm Killing vector $\xi$, called the R-symmetry 
vector field, or Reeb vector field. This vector field defines a foliation $\mathcal{F}_\xi$ of $Y_{2n+1}$ and is dual to the R-symmetry in the dual SCFTs in the cases above.
In local coordinates we may write 
\begin{align}
\xi = \partial_z\, , \qquad \eta = \diff z + P \, ,
\end{align}
where $\eta$ is the Killing one-form dual to $\xi$.
The metric on $Y_{2n+1}$ then has the form
\begin{align}\label{SEmetric}
\diff s^2_{2n+1}=\eta^2+ \diff s^2_{2n}\,,
\end{align}
where $\diff s^2_{2n}$ is a K\"ahler metric transverse to $\mathcal{F}_\xi$,  with transverse K\"ahler two-form $J$,
Ricci two-form $\rho$, and $\diff \eta =2J$. 
 Such geometries are known as Sasaki manifolds.
They can be considered to be
``off-shell" in the sense that when we impose the Einstein equations associated with \eqref{action1}, then we obtain
an ``on-shell" SE geometry. This is achieved if the transverse K\"ahler metric satisfies 
the Einstein condition
\begin{align}\label{einsteincond}
\rho=2(n+1)J\,,\qquad\Leftrightarrow \qquad R_{ij}=2(n+1) g_{ij}\, .
\end{align}

The real cone over the Sasaki manifold $Y_{2n+1}$ is a complex 
cone with conical  K\"ahler metric
\begin{align}\label{realcone}
\diff s^2_{2n+2} &= \diff r^2 + r^2 \diff s^2_{2n+1}~\,.
\end{align}
There is a natural compatible $SU(n+1)$ structure on this cone, with 
fundamental two-form $\mathcal{J}$ and holomorphic volume form $\Omega_{(n+1,0)}$, both of which are closed.
$\Omega_{(n+1,0)}$ is globally defined 
so the cone has zero first Chern class, i.e. it is CY. 
In addition, $\Omega_{(n+1,0)}$ has charge $n+1$ under the R-symmetry vector field $\xi$.
If we put the Sasaki geometry on-shell, then
the cone metric is both K\"ahler and Ricci-flat.
In the special cases of $n=2,3$ the $AdS$ solutions \eqref{typeiibsolsse}, \eqref{ansatzd11se} are associated with branes at the apex of the corresponding CY cone, as mentioned earlier.

If the orbits of the Killing vector are all circles then the SE manifold is called ``regular" or ``quasi-regular"
depending on whether the associated $U(1)$ action is free or not, respectively, and the R-symmetry is $U(1)$. In these cases the SE manifold
is a circle bundle over a K\"ahler-Einstein manifold or orbifold, respectively.
If the generic orbits are non-compact, which necessarily requires that the SE manifold has an 
additional Killing vector, then the SE manifold is called ``irregular" and the R-symmetry is $\mathbb{R}$. 

\subsection{Explicit solutions}
Many constructions of SE metrics are known, but explicit metrics are
rare. For simplicity we focus 
on $D=5$.
For $D=5$ the regular SE manifolds are circle bundles over four-dimensional K\"ahler-Einstein manifolds
and they are 
classified \cite{FRIEDRICHT1989Emod}. 
If simply connected they are given by $S^5$ and $T^{1,1}$, which are circle bundles
over $CP^2$ and $CP^1\times CP^1$, respectively, as well as circle bundles over del Pezzo surfaces $dP_k$, $3\leq k\leq 8$.
The metrics on $S^5$ and $T^{1,1}$ are homogeneous and explicitly known, while those associated with the del Pezzo's are not. 
The metric on $T^{1,1}$, which is topologically $S^2\times S^3$ and has $SU(2)^2\times U(1)$ isometry, was found in the supergravity literature in \cite{Romans:1984an}. 

The $Y^{p,q}$ are co-homogeneity one SE metrics on $S^2\times S^3$ with $SU(2)\times U(1)^2$ isometry \cite{Gauntlett:2004yd}; in fact they are the most general co-homogeneity one SE metrics in $D=5$ \cite{Conti:2006dg}. The $Y^{p,q}$ metrics include both quasi-regular and irregular SE metrics. 
Earlier, we recalled how the $Y^{p,q}$ metrics were originally discovered.
Another way to construct them is to consider the \emph{local} constructions of $D=4$ K\"ahler-Einstein metrics which are associated 
with the canonical  bundle over a K\"ahler-Einstein $CP^1$ base space \cite{bergery,Page:1985bq} and then 
add
an additional fibre direction. In fact this point of view can be generalized to construct SE metrics in higher dimensions by replacing the $CP^1$ base with a higher-dimensional K\"ahler-Einstein space, or a product of K\"ahler-Einstein spaces \cite{Gauntlett:2004hh}. 

The $L^{a,b,c}$ metrics \cite{Cvetic:2005ft} 
generalize the $Y^{p,q}$ metrics. They are explicit metrics on $S^2\times S^3$ but now with a reduced $U(1)^3$ isometry. Despite the reduced symmetry, it is still possible to construct these explicitly, essentially because they
have a Killing-Yano tensor \cite{Houri:2007xz}. Interestingly, the original construction of the $L^{a,b,c}$ metrics in \cite{Cvetic:2005ft}
was again rather indirect. Specifically, they were obtained by an analytic continuation
of some explicit Lorentzian rotating black hole solutions in $AdS_5$ spacetime \cite{Hawking:1998kw}.
The $L^{a,b,c}$ metrics are examples of toric SE metrics (in fact they are the most general toric metric with four rays
\cite{Martelli:2005wy}).

\subsection{Toric constructions}
\label{toricse}

Remarkable insights can be obtained for 
\emph{toric} SE metrics i.e. when
the isometry group of $Y_{2n+1}$ contains a maximal torus $\mathbb{T}^{n+1}=U(1)^{n+1}$. 
Importantly, it allows one to reformulate the volume extremization principle in algebraic terms and furthermore, it is the setting where most has been understood regarding the dual field theories.

The basic idea of toric geometry, in general, is to characterize a manifold (as well as generalizations, such as orbifolds, or cones \cite{Guillemin1994KaehlerSO,abreu2000kahler,Abreu:2001to,Burns2005KhlerMO}), 
in terms of loci where the torus action degenerates. Equivalently, 
a toric geometry consists of a torus fibration over a polytope 
with generic fibre being $\mathbb{T}^{n+1}$ in the interior of this polytope, while becoming sub-tori 
over the boundary of the polytope.

 For  Sasaki  manifolds $Y_{2n+1}$, an efficient approach is to view $Y_{2n+1}$ endowed with a $\mathbb{T}^{n+1}$ 
isometric action, as the link of the K\"ahler cone
 $C(Y_{2n+1})=\R_{>0}\times Y_{2n+1}$ with a toric action \cite{Martelli:2005tp}. In this case
 the polytope is a non-compact convex {polyhedral cone} 
  \begin{align}
{\cal C} = \{ \vec{y} \in \mathbb{R}^{n+1} ~ | ~  (\vec{y},\vec{v}_a) \geq 0 , \quad a=1,\dots, d\} \, ,
\end{align}
which may be obtained as the image of the moment map coordinates $\vec{y}$. 
Here the set $\{\vec{v}_a  \in  \mathbb{Z}^{n+1}, ~ a=1,\dots, d \geq n+1 \}$ of inward pointing primitive normals to the facets of  ${\cal C}$ is called the fan. 
The $\{\vec{v}_a\}$, often called the ``toric data", precisely determine which $U(1)\subset \mathbb{T}^{n+1}$ is degenerating on each facet.
The condition that the cone is CY, as discussed below \eqref{realcone}, translates into a
simple condition for the fan. 
A basis $\{\partial_{\varphi_i}\}$ for the $\mathbb{T}^{n+1}$ action can be chosen so that
the holomorphic volume form is only charged with respect to $\partial_{\varphi_1}$. In this case
the CY condition implies that  $\vec{v}_a  = (1,\vec{w}_a)$.

Topologically, the Sasaki manifold  $Y_{2n+1}$ may be reconstructed from the polyhedral cone  ${\cal C}$, by intersecting 
${\cal C}$ with a hyperplane, the ``Reeb hyperplane",
  \begin{align}
H(\vec{b}) = \{ \vec{y} \in \mathbb{R}^{n+1}, ~\vec{b}\in {\cal C}^* ~ | ~  (\vec{y},\vec{b} ) =\tfrac{1}{2}  \} \, .
\end{align}
This gives a compact, convex $n$-dimensional polytope 
  \begin{align}
P_n (\vec{b} )  = {\cal C} \cap H(\vec{b})  \subset \mathbb{R}^{n}\, ,  
\end{align}
whose vertices, generically,  do not lie in $\mathbb{Z}^{n}$.  In particular, $Y_{2n+1}$ can be presented as the toric fibration $\mathbb{T}^{n+1}\to Y_{2n+1}\to P_{n}(\vec{b})$.  As we will discuss further  below, the choice of vector $\vec{b} = (b_1,\dots,  b_{n+1}) \in {\cal C}^*$, where the latter denotes 
the dual cone to $\mathcal{C}$, corresponds to a choice of Reeb vector field $\xi=\sum_{i=1}^{n+1} b_i \partial_{\varphi_i}$, where $(y_i,\varphi_i)$ are Darboux coordinates on 
$C(Y_{2n+1})$ and $\{\partial_{\varphi_i}\}$ is a basis for an effective $\mathbb{T}^{n+1}$  action.  Remarkably, 
it turns out that every such  fan  $\{\vec{v}_a \}$, together with a
choice of  ``critical'' vector $\vec{b}_*$, that we discuss below, specify uniquely a SE metric \cite{Futaki:2006cc}. 

\subsection{Extremization principle}

\label{ext:section}

Inspired by physics considerations, 
it was found that SE geometry satisfies an interesting extremal problem \cite{Martelli:2005tp,Martelli:2006yb}. One of the virtues of this formalism is that one does not need the explicit SE metric in order to compute
various geometric quantities, including the volume of the SE manifold.
The general framework is to consider off-shell Sasaki geometry as follows.
We fix a complex cone $C(Y_{2n+1})=\R_{>0}\times Y_{2n+1}$ with holomorphic volume form $\Omega_{(n+1,0)}$, and holomorphic $U(1)^s$ action.  We then choose a ``trial'' holomorphic R-symmetry vector $\xi$ and demand that the holomorphic volume form has fixed charge $n+1$.
This choice of $\xi$ defines a foliation $\mathcal{F}_\xi$,
and fixing the charge of the holomorphic volume form 
furthermore fixes the basic cohomology class 
of the transverse K\"ahler metric as $[J]=[\rho]/(2n+2)\in  H^{1,1}_{{B}}(\mathcal{F}_\xi)$.
Crucially, we do not impose the transverse Einstein condition \eqref{einsteincond}, as this would immediately put us 
on-shell. 
Finally,  an on-shell geometry extremizes the Einstein-Hilbert action \eqref{action1}, which is simply proportional to the volume of the Sasaki manifold
\begin{align}\label{susyactvol}
\Vol  &=  \ \int_{Y_{2n+1}}\eta\wedge \tfrac{1}{n!}J^{n}\,.
 \end{align}
For fixed complex structure on the cone and fixed $\xi$, this is a 
holomorphic invariant, since the transverse K\"ahler class is fixed 
by the transverse complex structure via $[J]=[\rho]/(2n+2)\in  H^{1,1}_{{B}}(\mathcal{F}_\xi)$. Allowing the R-symmetry vector $\xi$ to vary then leads to $\Vol=\Vol(\xi)$ being a function, with SE metrics
 extremizing this function \cite{Martelli:2005tp,Martelli:2006yb}.

For the case of $n=2$, the extremal problem is associated with $AdS_5\times Y_5$ solutions of type IIB supergravity and
gives a precise geometric realization of $a$-maximization.
Similarly, when $n=3$, the above extremal problem is associated with supersymmetric solutions of $D=11$ supergravity
of the form $AdS_4\times Y_7$ and provides a geometric realization of F-extremization (which was found later).

In order to carry out the extremization one needs to be able to compute the volume as a function of $\xi$, and 
there are several ways to do this. A general formula was given in \cite{Martelli:2006yb} based on resolving the singularity of the cone over the Sasaki space and then applying localization.  
The volume can also be computed
as the limit of a certain index character which is constructed from the holomorphic functions on the cone \cite{Martelli:2006yb}. 
Another way appears in 
appendix E of \cite{Gauntlett:2019pqg} as a by-product of 
GK geometry.

In the toric case it is also possible to write down a very explicit formula in terms of the toric data \cite{Martelli:2005tp}. 
A basis of the $U(1)^{n+1}$ symmetry is chosen so that the holomorphic volume form is only charged under $\partial_{\varphi_1}$. 
The choice of an R-symmetry vector $\xi$ is specified by a vector $\vec{b}=(b_1,\dots,b_{n+1})$, with $b_1$ distinguished.
The toric geometry is specified by the troic data $\v_a$.
For example, when $n=2$, associated with $Y_5$, we have
  \begin{align}\label{sastoricvol}
	\text{Vol}(\vec{b};\{\vec{v}_a\})=\frac{\pi^3}{b_1}\sum_{a=1}^d \frac{[\vec{v}_{a-1},\vec{v}_a,\vec{v}_{a+1}]}{[\vec{b},\vec{v}_{a-1},\vec{v}_a] [\vec{b},\vec{v}_a,\vec{v}_{a+1}]}\, ,
\end{align}
where $[\cdot,\cdot,\cdot]$  denotes a $3\times 3$ determinant. 
Here the facets are ordered anti-clockwise around the polyhedral cone, and we cyclically identify $\v_{d+1}\equiv \v_1$, $\v_0\equiv \v_d$. We should also set $b_1=n+1$ and then extremize over the remaining components of $\vec{b}$.
An analogous expression when $n=4$, associated with $Y_7$, can be found in 
\cite{Gauntlett:2019roi}.

The volume function \eqref{susyactvol} 
 is a convex function of $\xi$, and 
 hence
there is always a unique critical point 
which is 
a minimum \cite{Martelli:2006yb}. 
Thus, if the SE metric exists,
the extremization identifies
the Reeb vector
and
the volume of the SE space.   
The critical vector $\vec{b}_*$ that extremizes the volume functional 
(\ref{sastoricvol}) obeys a set of rational
 equations and hence its components are, generically, 
algebraic numbers.
The same conclusion applies to the normalized volume $\text{Vol}(\vec{b}_*;\{\vec{v}_a\})/\pi^3$ of the SE metric. 
For quasi-regular examples, $\vec{b}_*$ and hence the normalized volume are rational numbers.
 In \cite{Futaki:2006cc} it was proven that for every such 
convex polyhedral cone $\mathcal{C}$, with critical 
vector $\vec{b}_*$, there is a unique 
SE metric on the base of the corresponding K\"ahler cone. 
This hence solves the general existence and uniqueness problem 
for toric SE metrics. 
 
Returning to the general (non-toric) case, 
by analysing the first variation of the volume function, the extremal equation can also be interpreted as setting the Futaki 
invariant of the transverse K\"ahler metric to zero \cite{Martelli:2006yb}. Thus, in the quasi-regular case the extremal problem can be 
understood as varying $\xi$ such that the transverse K\"ahler orbifold has zero Futaki invariant. The Futaki invariant was
a known obstruction to the existence of K\"ahler-Einstein metrics on K\"ahler manifolds and orbifolds, and so
the result of \cite{Martelli:2006yb} puts this into a more general context.

One can ask if there are further obstructions. Again using physical insight, two obstructions to
the existence of SE metrics on Sasaki manifolds were proven in \cite{Gauntlett:2006vf}.
The first, called the \emph{Bishop obstruction}, states that if the extremal volume is larger than that of the round sphere,
then there is no SE metric. The physical picture associated with this bound is as follows.
Consider
D3-branes sitting at the apex of a CY cone. By appropriately Higgsing the dual field theory and then integrating out the massive fields, one expects to 
flow to $\mathcal{N} = 4$ SYM theory. This is because the Higgsing corresponds to moving the D3-branes away from the singular point to a smooth point of the cone, where the near horizon geometry becomes $AdS_5 \times S^5$. Since the number of massless degrees of freedom should decrease under such a process, the $a$ central charge should decrease. Indeed this is the context of the $a$-theorem, 
later proven in
\cite{Komargodski:2011vj}. 
This implies 
the volume of the original SE metric associated with the CY cone, must necessarily
be smaller than that of the round $S^5$. A similar argument also holds for M2-branes. 

The second, called the \emph{Lichnerowicz obstruction}, states that if there are any holomorphic functions on the cone over
a Sasaki space with positive charge less than 1 under $\xi$, then there can be no SE metric. The physical intuition for this bound is simple.
Indeed, for the case of $AdS_5\times SE_5$ and $AdS_4\times SE_7$ solutions,
the existence of such a holomorphic function would translate into the dual SCFT as the statement that there are chiral
primaries which violate the unitary bound.

Subsequently, these ideas have been developed significantly in the mathematics literature, unifying with the notion of K-stability for K\"ahler-Einstein metrics 
and the Yau-Tian-Donaldson conjecture. Some key papers are 
\cite{Ross:2011big, Collins:2012dh}, culminating in \cite{Collins:2015qsb} 
which proves that a SE metric exists  on the base of a CY cone 
if and only if the cone is \emph{K-stable}, which is a purely 
algebro-geometric condition. The extremal problem  and Lichnerowicz obstruction 
described above are immediate corollaries. 
In some sense this work generalizes the above ideas: a
critical R-symmetry vector 
extremizes $\Vol$, but what if one has not extremized over a 
``large enough'' space, and so not actually found a critical point? 
Testing for 
K-stability involves looking at all possible ``degenerations'' of 
$C(Y_{2n+1})$, and checking $\diff \Vol\geq 0$ for those degenerations.  
This has also been given a direct interpretation in terms of AdS/CFT \cite{Collins:2016icw} and $a$-maximization.
Moreover, these results have been used 
to prove existence of further (non-toric) SE metrics e.g. new metrics on $S^2\times S^3$ (and connected sums of these), 
and on $S^5$.
Finally, we note that due to the existence theorem in \cite{Futaki:2006cc}
toric CY cones are always K-stable, and hence never obstructed.

\subsection{Dual field theories}

Before the discovery of $a$-maximization in 2003 \cite{Intriligator:2003jj}, there were two
main tools for identifying the $d=4$ SCFTs dual to $AdS_5\times SE_{5}$ solutions of type IIB supergravity. First, 
matching the global symmetries\footnote{For reasons of space,
we focus on flavour symmetries and don't discuss baryonic symmetries and baryonic operators associated with wrapped branes.} of the SCFT with the isometries of the $SE_5$ and, second, 
demanding that the (mesonic) vacuum moduli space of the SCFT should contain 
a branch that is precisely the CY cone over the $SE_5$, when viewed as a singular algebraic variety. 
The first examples that were identified consisted of ${\cal N}=1$ quiver gauge theories obtained as ``orbifold projections'' of 
${\cal N}=4$ SYM, with the SE dual given by $S^5/\Gamma$, where $\Gamma$ is a discrete subgroup of $SU(3)$
\cite{Douglas:1996sw}.  

Quiver gauge theories play a central role for SE more generally.
They
have a product gauge group $\Pi_i G_i$ with, 
in most cases, $G_i =SU(N_i)$  (or $U(N_i)$). 
In addition they have matter content 
consisting of chiral fields transforming in bi-fundamental representations of pairs of gauge groups. This data can be represented by oriented graphs, where the nodes correspond to vector multiplets transforming in the adjoint representation of $G_i$  and the oriented lines connecting the nodes are the bi-fundamentals. 

The construction of \cite{Douglas:1996sw} for $S^5/\Gamma$ was then
extended to a few non-orbifold examples in  \cite{Klebanov:1998hh,Morrison:1998cs}.
For the case of $T^{1,1}$, for which the metric was explicitly known, the large isometry group, $SU(2)^2\times U(1)$, was instrumental in determining the dual quiver 
\cite{Klebanov:1998hh}. A proposal for the suspended pinch point 
(SPP)
(toric) 
singularity
appeared in \cite{Morrison:1998cs}. However, since the corresponding explicit SE metric was not known, this proposal could not be conclusively checked until the later discovery 
of the $L^{a,b,c}$ SE metrics  \cite{Cvetic:2005ft}, which included the 
SPP 
as a special case \cite{Martelli:2005wy}.   
Another set of early constructions 
are the quiver gauge theories 
 associated with the toric del Pezzo singularities $dP_k$ with $k=1,2,3$ \cite{Feng:2000mi}. 
$dP_1$, $dP_2$ and $dP_3$ are all toric, with toric Calabi Yau cones, and hence are not obstructed.
$dP_3$ yields a regular SE metric, which is not explicitly known. $dP_1$ and $dP_2$ both give rise to irregular SE metrics with 
$dP_1$ corresponding to $Y^{2,1}$ [64], while the metric for $dP_2$, called $X^{2,1}$, is not explicitly known.
Quivers for the remaining del Pezzos $dP_k$, $k=4,\dots,8$ were found in \cite{Wijnholt:2002qz}.

For the general family of $Y^{p,q}$ metrics, the dual quiver gauge theories 
were obtained in \cite{Benvenuti:2004dy}, building on \cite{Martelli:2004wu,Bertolini:2004xf}.
In addition to utilising the global symmetries and an understanding of the corresponding toric CY singularities \cite{Martelli:2004wu},
a third and crucial tool was $a$-maximization \cite{Bertolini:2004xf,Benvenuti:2004dy}. In particular, this was used 
to show that the proposed quivers, which are defined in the UV, flow to SCFTs in the IR with a central charge precisely agreeing with the volume of the $Y^{p,q}$ manifolds via \eqref{acentse}.
A more precise check of the duality was further provided in 
\cite{Herzog:2006bu}, by explicitly matching a basis of fractional branes 
in the $C(Y^{p,q})$ geometries to the quivers. 

Similar strategies were used in \cite{Franco:2005sm} to identify the quiver gauge theories dual to the $L^{a,b,c}$ family of SE metrics. In parallel, increasingly efficient techniques, known as ``inverse algorithms'', were developed to extract the quiver gauge theories directly from toric CY singularities. This line of research culminated with the dimer models and brane tiling constructions \cite{Franco:2005rj,Hanany:2008fj,Franco:2017jeo}. 
Furthermore, an off-shell agreement between $a$-maximization and volume minimization for toric $D=5$ SE was demonstrated in \cite{Butti:2005vn} (and later extended to the non-toric case in \cite{Eager:2010yu}). 
The structure of the superconformal multiplets was obtained from the toric SE geometry in
\cite{Eager:2012hx} and this was also successfully matched with a computation of the
single-trace superconformal index in the quiver gauge theory.
In light of all these developments, combined with the fact that the existence of SE metrics in the toric class are not obstructed \cite{Futaki:2006cc}, the AdS/CFT correspondence 
for $AdS_5\times SE_5$ solutions in the toric case can be considered to be rather well understood 
(at leading order in the large $N$ limit). 

Much less is understood for the case of non-toric $SE_5$ and the general goal of identifying the dual SCFTs
remains largely open. 
There are by now many existence 
theorems for (non-toric) SE metrics, starting with the early 
work of \cite{Boyer:2000pg, BoyerCharlesP.2001NEMi} which constructed the 
very first examples of quasi-regular (but not regular) SE metrics. 
Such constructions were subsequently generalized extensively, 
with many classes of infinite families of non-toric, quasi-regular 
SE metrics. Furthermore, using the more recent 
results on K-stability  \cite{Collins:2015qsb} (mentioned in section \ref{ext:section}), 
infinite classes of non-toric, irregular SE metrics were constructed in \cite{suss2021}, 
where the latter reference also proves that no irregular SE metric exists with the topology of $S^5$. In general, the dual field theories for these SE manifolds
remain obscure. 
However, some constructions have been reported in \cite{Fazzi:2019gvt} and one can 
hope that further progress in this direction can be made.
 
The analogous story for the $d=3$ ${\cal N}=2$ SCFTs dual to $AdS_4\times SE_7$ solutions is 
more complicated and 
much less understood.
Some initial proposals were 
put forward in 1999 based on the Kaluza-Klein spectra of various coset manifolds \cite{Fabbri:1999hw,Ceresole:1999zg}, but key ideas about the physics of M2-branes were still missing. 
It took almost ten more years, building on various important work including \cite{Schwarz:2004yj,Bagger:2006sk,Hosomichi:2008jd},
before the 
$d=3$ Chern-Simons-matter quiver gauge theory (ABJM) dual to $AdS_4\times S^7$ was found \cite{Aharony:2008ug}. 

The geometric extremization results of \cite{Martelli:2005tp,Martelli:2006yb} indicated the existence of an extremization principle for  $d=3$ ${\cal N}=2$ SCFTs. This was later formulated
as the principle of F-extremization, whereby one can compute the partition function of the dual SCFT when placed on the round $S^3$
by extremizing over the space of possible R-symmetries \cite{Jafferis:2010un}. 
The connection with SE geometry is then via \eqref{fexps3}.
This field theory principle has again been an important tool in identifying the dual SCFTs
along with the expectation that the CY four-fold singularity should arise as a branch of the moduli space of vacua  \cite{Martelli:2008si,Hanany:2008fj,Jafferis:2008qz}.
However, currently it is only known how to carry out F-extremization for a very limited class of field theories, dubbed ``non-chiral'' 
\cite{Martelli:2011qj,Jafferis:2011zi},
 leaving a large gap in our understanding of the AdS$_4$/CFT$_3$ correspondence in this context. Nonetheless, various Chern-Simons-matter quivers have been proposed in the toric case \cite{Martelli:2008si,Hanany:2008fj,Jafferis:2008qz,Benini:2009qs,Benini:2011cma,Closset:2012ep,Amariti:2012tj} as well as in the non-toric case  \cite{Martelli:2009ga,Jafferis:2009th}. 

\section{GK geometry}

\subsection{GK geometry}

GK geometry \cite{Gauntlett:2007ts}
is defined on an odd-dimensional manifold, $Y_{2n+1}$, with $n\ge 3$ 
and consists of a metric, a scalar function $B$ and a closed two-form $F$.
Writing $F=\diff A$, these quantities satisfy equations of motion that can be derived from the action
\begin{align}\label{action}
S 
 =  \int_{Y_{2n+1}} \ex^{(1-n)B}\left[R_{2n+1} - \frac{2n}{(n-2)^2}+\frac{n(2n-3)}{2}(\diff B)^2 + \frac{1}{4}\ex^{2B}F^2\right]\vol_{2n+1}~.
\end{align}
The geometry also admits certain Killing spinors $\epsilon$ satisfying
\begin{align}
\Big(\nabla_a+\frac{\ii}{2}\gamma_a+\frac{1}{8}\ex^BF_{bc}\gamma_a{}^{bc}\Big)\epsilon&=0\,,\nn
\Big(\gamma^a\nabla_a B+\ii\frac{2(n-1)}{n-2}+\frac{1}{2}\ex^BF_{ab}\gamma^{ab}\Big)\epsilon&=0\,.
\end{align}

The 
cases relevant for physics are $Y_7$ and $Y_9$, when $n=3$ and $n=4$. 
When $n=3$ we obtain  supersymmetric $AdS_3\times Y_7$ solutions of type IIB supergravity of the form \cite{Kim:2005ez}
\begin{align}\label{typeiibsols}
\diff s^2_{10} &= L^2 \ex^{-B/2}\left(\diff s^2({AdS_3}) + \diff s^2({Y_7})\right)~,\nn
F_5 &= -L^4\left(\vol_{{AdS}_3}\wedge F + *_7 F\right)~,
\end{align}
where $\diff s^2({{AdS}_3})$ has unit radius, and $L$ is fixed by flux quantization.
These solutions are dual to $d=2$ SCFTs with
$\mathcal{N}=(0,2)$ supersymmetry which necessarily have an R-symmetry.  A key observable to be computed from such solutions
is the central charge $c$ which can be obtained by computing the three-dimensional Newton constant.
An important way (but not the only way) 
in which these solutions arise is by considering a $d=4$ $\mathcal{N}=1$ SCFT that is dual to an
$AdS_5\times SE_5$ solution. If one places this on $\mathbb{R}^{1,1}\times\Sigma$, with $\Sigma$ a Riemann surface
or a spindle, in such a way to preserve supersymmetry, generically one expects an RG flow to a 
$d=2$ SCFT. Thus, associated with this picture, we expect rich classes of solutions where $Y_7$ is a fibration of the form
$SE_5\to Y_7\to\Sigma$ and these would arise as the near horizon limit of black strings in $AdS_5\times SE_5$.

Similarly, when $n=4$ we obtain supersymmetric $AdS_2\times Y_9$
solutions of $D=11$ supergravity
of the form \cite{Kim:2006qu}
\begin{align}\label{ansatzd11}
\diff s^2_{11} &= L^2 \ex^{-2B/3}\left(\diff s^2({{AdS}_2}) + \diff s^2(Y_{9})\right)~,\nn
G_4 &= L^3\vol_{{AdS}_2}\wedge F ~,
\end{align}
where $\diff s^2({{AdS}_2})$ has unit radius and $L$ is fixed by flux quantization. These solutions are dual (formally) to $d=1$ SCFTs with $\mathcal{N}=2$
supersymmetry with an R-symmetry. They are of particular interest as they arise as the near horizon limit of supersymmetric black holes.
In this setting a key observable to be computed is the black hole entropy $S_{\mathrm{BH}}$ which can be obtained by computing
the two-dimensional Newton constant.
An important way in which these solutions arise is by considering a $d=3$ $\mathcal{N}=2$ SCFT that is dual to an
$AdS_4\times SE_7$ solution. Placing this field theory on $\mathbb{R}^{}\times\Sigma$ and preserving supersymmetry, generically one  
expects rich classes of solutions with $Y_9$ a fibration of the form $SE_7\to Y_9\to\Sigma$, which would arise as the near horizon limit of black holes in $AdS_4\times SE_7$. In this setting, following \cite{Benini:2015eyy},
an important goal is to recover the black hole entropy  $S_{\mathrm{BH}}$, by carrying out a field
theory computation associated with the dual $d=3$ $\mathcal{N}=2$ SCFT compactified on $\Sigma$.

The Killing spinor equations imply that the metric on $Y_{2n+1}$ has a unit norm Killing vector $\xi$, called the R-symmetry 
vector field, which defines a foliation $\mathcal{F}_\xi$ of $Y_{2n+1}$. In local coordinates we may write 
\begin{align}
\xi = \frac{1}{c}\partial_z\, , \qquad \eta = c(\diff z + P)\, ,
\end{align}
where $c\equiv (n-2)/2$ and $\eta$ is the Killing one-form dual to $\xi$.
The metric on $Y_{2n+1}$ then has the form
\begin{align}\label{GKmetric}
\diff s^2_{2n+1}=\eta^2+\ex^B \diff s^2_{2n}\,,
\end{align}
where $\diff s^2_{2n}$ is a K\"ahler metric transverse to $\mathcal{F}_\xi$. This K\"ahler metric, with transverse K\"ahler two-form $J$,
Ricci two-form $\rho$ and Ricci scalar $R$, fixes the remaining geometry via
\begin{align}
\rho=\frac{1}{c}\diff \eta\,,\qquad\qquad
\ex^B=\frac{c^2}{2}R\,,\qquad\qquad
F=-\frac{1}{c}J+\diff \left(\ex^{-B}\eta\right)\, ,
\end{align}
which imply positive scalar curvature, $R>0$. 
These off-shell ``supersymmetric geometries" become on-shell GK geometries satisfying the
equations of motion coming from \eqref{action},
provided 
the transverse K\"ahler metric satisfies 
\begin{align}\label{boxR}
\Box R &=\frac{1}{2}R^2 - R_{ij}R^{ij}~.
\end{align}
Clearly there are strong similarities with the SE case, with the off-shell supersymmetric geometry being the analogue of Sasaki geometry and
\eqref{boxR} the analogue of the Einstein condition on the transverse metric \eqref{einsteincond}. 

Another feature in common with Sasakian geometry is that the real cone over $Y_{2n+1}$ is a complex 
cone with conical (non-K\"ahler) metric
\begin{align}
\diff s^2_{2n+2} &= \diff r^2 + r^2 \diff s^2_{2n+1}~\,.
\end{align}
There is a natural compatible $SU(n+1)$ structure on this cone, with 
fundamental two-form $\mathcal{J}$ and holomorphic volume form $\Omega_{(n+1,0)}$.
The two-form $\mathcal{J}$ is not closed and so unlike the Sasaki case, there is no natural symplectic 
structure on the cone. However, $\Omega_{(n+1,0)}$ is both globally defined and conformally closed,
so the cone has zero first Chern class, as in the Sasaki case. In addition, $\Omega_{(n+1,0)}$ has charge $1/c$ under the R-symmetry vector field $\xi$.

The final ingredient is a flux quantization condition.
For $n=3$ and $n=4$ we obtain $AdS_3\times Y_7$ 
and $AdS_2\times Y_9$ solutions of type IIB and $D=11$ supergravity, as in \eqref{typeiibsols},\eqref{ansatzd11}
and we must impose flux quantization for cycles of co-dimension two. 
This naturally generalizes to all $n\geq 3$. Specifically, if $\Sigma_A$ are a basis for $H_{2n-1}(Y_{2n+1},\mathbb{Z})$
we impose
\begin{align}\label{fluxqc}
\int_{\Sigma_A}\left[\eta \wedge \rho\wedge \tfrac{1}{(n-2)!}J^{n-2}  + \frac{c}{2}*_{2n}\diff R\right]=\nu_{n} N_A\,,
\end{align}
where $N_A\in\mathbb{Z}$ and the non-zero, real constant $\nu_n$ is explicitly fixed only for $n=3$ and $n=4$:
$\nu_3={2(2\pi \ell_s)^4g_sL^{-4}}$, where $\ell_s$ is the string length, and $g_s$ is the string coupling
and
$\nu_4=(2\pi \ell_p)^6L^{-6}$,
where $\ell_p$ is the eleven-dimensional Planck length.

As in the SE case, if the orbits of the R-symmetry are all circles we have regular or quasi-regular GK geometry, depending on whether the action is free or not. If the R-symmetry orbits are generically $\mathbb{R}$ then the GK geometry is irregular.

\subsection{Explicit solutions}
Various explicit examples of GK geometry are known
\cite{Gauntlett:2006af,Gauntlett:2006qw,Gauntlett:2006ns,Donos:2008ug,Gauntlett:2018dpc,Couzens:2019mkh}, and 
these have played an important role in guiding further developments. 
One fruitful strategy is to mimic approaches for constructing explicit SE metrics. For example, one can consider 
constructions of \emph{local} K\"ahler metrics of precisely the type arising in SE constructions \cite{Gauntlett:2004hh} and then instead of imposing
the transverse Einstein condition \eqref{einsteincond}, one imposes the ``box $R$" condition \eqref{boxR}. One then adds an additional
fibre direction as in \eqref{GKmetric} and demands that the resulting metric is  globally defined on a regular manifold $Y_{2n+1}$.

Some of these explicit solutions are of the fibred form $SE_{2n-1}\to Y_{2n+1}\to\Sigma$, where $\Sigma$ is two-dimensional.
For $n=3,4$ these can be interpreted as D3-branes or M2-branes wrapping two-cycles in $CY_4$, $CY_5$ spaces,
with the normal space a $CY_3$ or $CY_4$ cone, respectively. 
There are constructions where $\Sigma$ is a Riemann surface of genus $g$ with a constant curvature metric and supersymmetry is preserved with a ``topological twist" \cite{Witten:1988ze}, connecting with the large class of wrapped brane solutions arising in many different settings, starting with \cite{Maldacena:2000mw}. More recently, it has also been understood \cite{Ferrero:2020laf} that there are constructions in which $\Sigma$ is a \emph{spindle}, i.e. a two-sphere with an azimuthal symmetry and orbifold singularities at the poles, and furthermore that supersymmetry can be preserved in two distinct ways, the ``twist" and the ``anti-twist"
\cite{Ferrero:2021etw}. Interestingly, the total GK spaces $Y_{2n+1}$ are regular, despite the presence of the orbifold singularities
on the spindle. For the case of $n=4$, it has also been shown that the $AdS_2\times Y_9$ solutions arise as the near horizon limit of 
accelerating black holes in $AdS_4$ \cite{Ferrero:2020twa}.
 
The brane interpretation of the above fibred class of GK geometries has been key to identifying the dual field theories in many 
cases \cite{Gauntlett:2018dpc,Hosseini:2019use}. For example, consider the $d=4$ quiver gauge theories dual to $AdS_5\times SE_5$, with toric $SE_5$. When these are
reduced on $\Sigma$ they generically flow to a $d=2$ SCFT in the IR and, for example, the central charge can be computed using
$c$-extremization \cite{Benini:2012cz,Benini:2013cda}. These are expected to be dual to 
$AdS_3\times Y_7$ solutions with $SE_{5}\to Y_{7}\to\Sigma$. Strong evidence for this general picture is obtained by matching
the central charge (and some other BPS observables) computed on the field theory side with the result obtained from the explicit GK solution. However, 
there are some examples which are not understood, including the case of $\Sigma=T^2$ \cite{Couzens:2018wnk}.

\subsection{Extremization principle}

Inspired by physics, 
it was found that GK geometry satisfies an 
extremal problem \cite{Couzens:2018wnk}, analogous to that of SE geometry. 
When $n=3$, the associated $AdS_3\times Y_7$ solutions
\eqref{typeiibsols} are dual to $d=2$ $\mathcal{N}=(0,2)$ SCFTs. 
The $c$-extremization principle \cite{Benini:2012cz,Benini:2013cda} for these SCFTs suggested
there could be an extremization principle for $Y_7$, but the principle found in \cite{Couzens:2018wnk} is applicable to all GK geometry. Importantly, the extremization principle for $Y_7$ and $Y_9$ allows one to compute the central charge of the dual field
theory and the entropy of black holes (as well as some other BPS observables) without having an explicit solution, just knowing some global topological data.

The formalism is similar to that of SE geometry, but a little more involved.
We start by considering an off-shell supersymmetric geometry as follows.
We fix a complex cone $C(Y_{2n+1})=\R_{>0}\times Y_{2n+1}$ with holomorphic volume form $\Omega_{(n+1,0)}$, and holomorphic $U(1)^s$ action. 
We then choose a trial holomorphic R-symmetry vector $\xi$ and demand that the holomorphic volume form has fixed charge $1/c$.
This choice of $\xi$ defines a foliation $\mathcal{F}_\xi$, and  we then further choose a transverse K\"ahler metric with 
basic cohomology class $[J]\in H^{1,1}_{{B}}(\mathcal{F}_\xi)$. 
Crucially, we do not impose the condition \eqref{boxR}, as this would put us 
on-shell. However, in order to impose the flux quantization condition (\ref{fluxqc}) we impose that the integral of \eqref{boxR} is satisfied. Specifically, we impose
the topological constraint
\begin{align}\label{constraint}
\ \int_{Y_{2n+1}}\eta\wedge \rho^2\wedge \tfrac{1}{(n-2)!}J^{n-2}=0\,,
 \end{align}
as well as the flux quantization conditions
\begin{align}\label{quantize}
\int_{\Sigma_A}\eta\wedge \rho\wedge \tfrac{1}{(n-2)!}J^{n-2}&\ = \nu_n N_A\,,
\end{align}
with the basis of co-dimension two cycles $\{{\Sigma_A}\}~$ all tangent to $\xi$.
Finally,  an on-shell geometry, with properly quantized flux, extremizes the supersymmetric action
\begin{align}\label{susyact}
\Ssusy  \ &=  \ \int_{Y_{2n+1}}\eta\wedge \rho\wedge \tfrac{1}{(n-1)!}J^{n-1}\,.
 \end{align}
For a given $\xi$, the quantities 
\eqref{constraint}--\eqref{susyact} 
just depend on the basic cohomology class $[J]\in H^{1,1}_{{B}}(\mathcal{F}_\xi)$, and not on $J$ itself \cite{Couzens:2018wnk}. 
In contrast to the Sasakian case, now it is the flux numbers 
$N_A$ which effectively determine the transverse K\"ahler class, 
and a GK geometry 
with quantized flux 
is necessarily an extremal point of the action, as a function of $\xi$ for fixed flux numbers. As discussed in \cite{Couzens:2018wnk}, \cite{Gauntlett:2018dpc}, for a given extremal point there may be obstructions to the existence of a 
corresponding GK supergravity solution (i.e. satisfying  (\ref{boxR})). 
Further exploring obstructions with the goal of determining sufficient conditions for the existence
of a GK geometry is an important open problem. 

For the case of $n=3$, the extremal problem is associated with $AdS_3\times Y_7$ type IIB solutions.
Remarkably, the value of the on-shell action determines the central charge, $c$, of the dual field theory. 
Specifically, defining an ``off-shell central charge", $\cZ$, via
\begin{align}\label{cS}
\cZ  \equiv  \frac{3L^8}{(2\pi)^6g_s^2\ell_s^8} \Ssusy=\frac{12 (2\pi)^2}{\nu_3^2} \Ssusy~,
\end{align}
where $\Ssusy$ is the supersymmetric action (\ref{susyact}) with $n=3$, then we have
\begin{align}\label{cS2z}
\cZ  |_\mathrm{on-shell}  =  c~.
\end{align}
This provides a precise geometric realization of $c$-extremization.

Similarly, when $n=4$ the above extremal problem is associated with supersymmetric solutions of $D=11$ supergravity
of the form $AdS_2\times Y_9$.
For this case we can define an ``off-shell entropy", $\mathscr{S}$, via
\begin{align}\label{cS2}
\mathscr{S}\equiv\frac{4\pi L^9}{(2\pi)^8\ell_p^9}\, \Ssusy~,
\end{align}
where $\Ssusy$ is the supersymmetric action (\ref{susyact}) with $n=4$.
In the case that the $D=11$ solution arises as the near horizon limit of a supersymmetric black hole, it
is expected that $\mathscr{S}|_\mathrm{on-shell}$ is the entropy of the black hole \cite{Couzens:2018wnk}.
More generally, it is expected that $\mathscr{S}|_\mathrm{on-shell}$ is the logarithm of a supersymmetric partition function
of the dual quantum mechanical theory \cite{Couzens:2018wnk}. For the sub-class of solutions for which $Y_9$ is of the fibred form
$SE_{7}\to Y_{9}\to\Sigma$, with $\Sigma$ a Riemann surface, \cite{Gauntlett:2019roi, Hosseini:2019ddy} established a
connection with the field theory notion of $\mathcal{I}$-extremization, thus providing
a state counting interpretation of the entropy of infinite classes of supersymmetric
$AdS_4$ black hole solutions.
When $\Sigma$ is a spindle a computation of the large $N$ spindle index was obtained in \cite{Colombo:2024mts} and shown to agree with the entropy function associated with accelerating black holes \cite{Cassani:2021dwa}.

\subsection{Toric constructions}
The extremization principle for GK geometry can be further
developed using toric geometry. 
Interestingly, the relevant toric geometry has some novel features and while the formalism has been extensively checked using explicit solutions, there are aspects that still need to be put on a rigorous mathematical foundation.

We begin with a K\"ahler cone metric on a complex cone $X_{2r+1}$ and we let $\xi$ be an R-symmetry vector.
In the Sasaki setting, the R-symmetry vector is the Reeb vector associated with a transverse K\"ahler metric with K\"ahler form 
$\omega=\omega_\text{Sasakian}$. For GK geometry we are
interested in a more general class of transverse K\"ahler metrics $\omega$ and in particular we are interested in
the \emph{master volume} defined by
\begin{align}\label{mastvol}
\mathcal{V}(\xi,[\omega])=\int_{X_{2r+1}}\eta\wedge \frac{1}{r!}\omega^r\,,
\end{align}
which is a function of $\xi$ and the transverse K\"ahler class $[\omega] \in H^{1,1}_B(\mathcal{F}_\xi)$.

If $X_{2r+1}$ is toric there is a holomorphic $U(1)^{r+1}$ symmetry and the choice of an R-symmetry vector  field $\xi$ is specified by a vector $\vec{b}=(b_1,\dots,b_{r+1})$. A basis of the $U(1)^{r+1}$, $\{\partial_{\varphi_i}\}$,
is chosen so that the holomorphic volume form is only charged under $\partial_{\varphi_1}$. The toric geometry is specified by
the vectors $\{\v_a=(1,\w_a)\in\Z^{r+1}\mid a=1,\ldots,d\}$, similar to section \ref{toricse},
which in particular singles out the component $b_1$ as being special. 
It should be emphasized that unlike ordinary toric geometry, for the applications to GK geometry
we do not demand that the $\vec{v}_a$ define a convex polytope and so we are considering a novel kind of ``non-convex" toric geometry \cite{Couzens:2018wnk,Gauntlett:2018dpc}. The class $[\omega]$ 
can be parametrized by $d$ real parameters $\lambda_a$, of which only $d-r$ are independent.
In the case of $r=2$ a very explicit expression for the master volume of $X_5$ is given by \cite{Gauntlett:2018dpc}
\begin{align}\label{toricmaster}
\mathcal{V}(\vec{b};\{\lambda_a\};\{\vec{v}_a\})  =  \frac{(2\pi)^3}{2}\sum_{a=1}^d \lambda_a \frac{\lambda_{a-1}[\v_a,\v_{a+1},\vec{b}] - \lambda_a [\v_{a-1},\v_{a+1},\vec{b}]+\lambda_{a+1}[\v_{a-1},\v_a,\vec{b}]}{[\v_{a-1},\v_a,\vec{b}][\v_a,\v_{a+1},\vec{b}]}~,
\end{align}
where $[\cdot,\cdot,\cdot]$  denotes a $3\times 3$ determinant. Here the facets are ordered anti-clockwise around the polyhedral cone, 
and we cyclically identify $\v_{d+1}\equiv \v_1$, $\v_0\equiv \v_d$, and similarly $\lambda_{d+1}\equiv \lambda_1$, 
$\lambda_0\equiv \lambda_d$. An analogous, but more complicated explicit expression for the case of $r=3$, associated with $X_7$, is given in 
\cite{Gauntlett:2019roi}. 
In the special case in which 
\begin{align}
\label{lambdaSas}
\lambda_a  =  -\frac{1}{2b_1}\,, \qquad a=1,\dots d\,,
\end{align}
we recover the Sasakian  K\"ahler class, and the master volume (\ref{mastvol}) reduces to the Sasakian volume \eqref{sastoricvol}.
In general, the master volume is invariant under
\begin{align}\label{lamgt}
\lambda_a\ \to\ \lambda_a+\sum_{i=1}^{r+1}\gamma_i(v_a^i b_1-b_i)\,,
\end{align}
for arbitrary constants $\gamma_i$, which can be viewed as ``gauge" transformations.

We can employ this toric data in several different ways in GK geometry. The first and most obvious way, is when the toric $X_{2r+1}$ is itself the GK geometry. In this case the extremization problem can be recast purely in terms of the master volume $\mathcal{V}(\vec{b};\{\lambda_a\};\{\vec{v}_a\})$. For example, if $X_7$ is a toric GK geometry, associated with an $AdS_3\times X_7$ solution, we have
\begin{align}
0 & =   \sum_{a,b=1}^d \frac{\partial^2 \mathcal{V}}{\partial \lambda_a\partial \lambda_b}\,, \quad 
\frac{2(2\pi \ell_s)^4g_s}{L^4}N_a  =  \frac{1}{2\pi}\sum_{b=1}^d \frac{\partial^2 \mathcal{V}}{\partial \lambda_a\partial \lambda_b}~, \quad \Ssusy   =   -\sum_{a=1}^d \frac{\partial \mathcal{V}}{\partial \lambda_a}~.\label{ssusytoricseven}
\end{align}
The first and second expressions are the constraint equation \eqref{constraint}
and the flux quantization condition \eqref{quantize}, with $N_a\in\mathbb{Z}$ (not all
of which are independent). One should set $b_1=2$ and then extremize $\Ssusy$, or equivalently the off-shell central charge $\cZ$ in \eqref{cS},
 with respect to the $b_2,b_3,b_4$ and the $d-3$ independent $\lambda_a$. The on-shell expression for $\cZ$
 then gives (assuming the GK geometry actually exists) the central charge of the dual
SCFT. This formalism was checked against explicit solutions in \cite{Gauntlett:2019roi}, but the more significant point is that 
it applies
to solutions that are unlikely ever to be found in explicit form. 
All of these toric $X_7$ GK examples have rational R-charges and central charges, 
as one expects from $c$-extremization on the field theory side. Indeed, it was conjectured
this applies more generally\footnote{Note that this is special to $Y_7$; an example of irregular GK $Y_9$ is discussed in appendix C of  \cite{Gauntlett:2019roi}.} and
that $Y_7$ GK geometry is always in the regular or quasi-regular class \cite{Gauntlett:2018dpc}.
A similar investigation for toric $X_9$, associated with $AdS_2\times X_9$ solutions, has not yet been carried out.

The second way is if the toric $X_{2r+1}$ arises as the fibre of a GK geometry. Specifically we can consider GK geometry $Y_{2r+2k+1}$ of the form $X_{2r+1}\to Y_{2r+2k+1}\to B_{2k}$ where $B_{2k}$ is a K\"ahler geometry.
This has been explored in \cite{Gauntlett:2019pqg}, assuming that the R-symmetry vector $\xi$ is tangent to the toric fibres.
From a physical perspective, the most interesting case is when $k=1$ with
$X_{2r+1}\to Y_{2r+3}\to \Sigma$, where $\Sigma$ is two-dimensional. 
This includes GK geometry of the form $X_{5}\to Y_{7}\to \Sigma$ and 
$X_{7}\to Y_{9}\to \Sigma$, associated with $AdS_3\times Y_7$ and $AdS_2\times Y_9$ geometries, and one can view them as arising from RG flows from $AdS_5\times X_5$ and $AdS_4\times X_7$, where $X_5$ and $X_7$ are SE, respectively.
In fact 
this set-up falls into two further different cases: $\Sigma$ could either be a Riemann surface of genus $g$, $\Sigma_g$, or  a spindle.

We first illustrate how the extremization procedure works for the first case in the setting
$X_{5}\to Y_{7}\to \Sigma_g$, with $X_5$ toric and admitting
an isometric $U(1)^3$ action. The extremization problem can now be phrased in terms of the master volume of $X_5$.
A new ingredient is that the fibration structure needs to be specified by topological data, which is captured by the vector $(n_1,n_2,n_3)\in\mathbb{Z}^3$ (any $U(1)^3$ bundle over a Riemann surface is classified by three Chern numbers). 
Consistency requires $n_1=2(2-g)$, which corresponds to supersymmetry being preserved via a topological twist \cite{Witten:1988ze,Maldacena:2000mw}.
The constraint equation \eqref{constraint} now reads
\bea\label{constraintformula} 
0 &=& A \sum_{a,b=1}^d \frac{\partial^2\mathcal{V}}{\partial\lambda_a\partial\lambda_b} - 2\pi n_1 \sum_{a=1}^d\frac{\partial \mathcal{V}}{\partial\lambda_a} + 2\pi b_1 \sum_{a=1}^d \sum_{i=1}^3 n_i\frac{\partial^2 \mathcal{V}}{\partial \lambda_a\partial b_i}~,
\eea
where $A$ is a K\"ahler class parameter for $\Sigma_g$. The flux quantization conditions coming from \eqref{quantize} are equivalent to
\bea\label{Nformula}
\frac{2(2\pi \ell_s)^4 g_s}{L^4}N &=& -\sum_{a=1}^d \frac{\partial \mathcal{V}}{\partial\lambda_a}~,\\
\label{Maformula}
\frac{2(2\pi \ell_s)^4 g_s}{L^4}M_a &=& \frac{A}{2\pi}\sum_{b=1}^d\frac{\partial^2\mathcal{V}}{\partial\lambda_a\partial\lambda_b}+b_1\sum_{i=1}^3 n_i\frac{\partial^2 \mathcal{V}}{\partial\lambda_a\partial b_i} ~,
\eea
with $N,M_a\in\mathbb{Z}$, which are associated with the fibre five-cycle $X_5$ or with a basis of three-cycles on $X_5$ fibred over the Riemann surface, respectively.
The supersymmetric action has the form
\bea\label{Ssusyformula}
\Ssusy  &=& -A \sum_{a=1}^d \frac{\partial\mathcal{V}}{\partial \lambda_a} - 2\pi b_1 \sum_{i=1}^3 n_i \frac{\partial\mathcal{V}}{\partial b_i}\,.
\eea
This needs to be extremized after setting $b_1=2$ (after taking any derivatives with respect to $\vec{b}$).
Extremizing this action or, equivalently, the off-shell central charge $\cZ$ in \eqref{cS} with respect to the
independent K\"ahler class parameters and R-symmetry data, one obtains
the on-shell central charge $c  =  \left.\cZ\right|_{\mathrm{on-shell}}$.

This formalism has been checked using infinite classes of explicit supergravity solutions, with exact agreement \cite{Gauntlett:2018dpc}.
Furthermore, precise agreement with field theory has also been found. On the field theory side 
one starts with the quiver gauge theory that is dual
to an $AdS_5\times SE_5$ solution, with toric $SE_5$. One places the field theory on a Riemann surface and, assuming
one flows to a $d=2$ SCFT in the IR, one computes the central charge using $c$-extremization.
Remarkably, there is actually an off-shell agreement between the 
geometric and the field theory versions of $c$-extremization \cite{Gauntlett:2018dpc}, which was 
proven in general in \cite{Hosseini:2019use}.
This provides a general exact result for AdS/CFT. We highlight, though, that a key assumption 
on the geometry side is that the GK geometries $Y_7$ actually exist i.e. they are not obstructed (an assumption that also applies to the discussion below). A conjecture for sufficient conditions for existence was given in \cite{Gauntlett:2018dpc}, based on
the physical constraints that the central charge and the R-charges of certain BPS operators should all be positive.

An analogous investigation has been carried out for $AdS_2\times Y_9$ solutions with $X_{7}\to Y_{9}\to \Sigma_g$ in
\cite{Gauntlett:2019roi,Hosseini:2019ddy}, and the off-shell entropy function \eqref{cS2} can be obtained in a similar way.
The resulting on-shell entropy has been explicitly checked against known solutions, as well as being applied to cases 
where solutions are not known. 
Moreover, the on-shell entropy has also been shown to precisely agree with the field theory computation
of the twisted index using $\mathcal{I}$-extremization, for certain cases where the latter has been computed, as in \cite{Hosseini:2016tor,Hosseini:2016ume}.
Together, this provides a microstate counting interpretation for the entropy of infinite classes of black holes,
substantially extending the initial examples of \cite{Benini:2015eyy,Azzurli:2017kxo}.

The final case we highlight is GK geometry of the form
$X_{2r+1}\to Y_{2r+3}\to \Sigma$, where now $\Sigma$ is a spindle. 
In the case of $AdS_3\times Y_7$ solutions, these are associated with
SCFTs arising from D3-branes wrapping spindles \cite{Ferrero:2020laf,Hosseini:2021fge,Boido:2021szx,Ferrero:2021etw}. In the case of $AdS_2\times Y_9$ solutions, they arise as the near horizon limit of accelerating black holes \cite{Ferrero:2020twa,Cassani:2021dwa,Ferrero:2021ovq,Ferrero:2021etw,Couzens:2021rlk}.
A key difference from the case when $\Sigma$ is a Riemann surface is that the
R-symmetry vector can now be tangent to both the fibre and the spindle.\footnote{For genus $g=0$ Riemann surfaces, one can also allow the R-symmetry to rotate the $S^2$. However, generically one finds, after extremization, that this is not possible.}
An important general result, with the fibre $X_{2r+1}$ not necessarily toric,
is that the off-shell supersymmetric action can be written in the form
\cite{Boido:2022mbe}
 \begin{align}
 \label{ssusytomaster}
	S_{\text{SUSY}}	= \frac{2\pi b_1}{b_0} \left(
		\mathcal{V}_{2r+1}^+ - \mathcal{V}_{2r+1}^- \right) \,,
\end{align}
where $\mathcal{V}_{2r+1}^\pm$
are the master volumes of the fibres $X_\pm$, over the north and south poles of the spindle, and $b_0$ is the component of the 
R-symmetry vector tangent to the spindle.
More precisely, if $\xi_\pm$ is the orthogonal projection of the R-symmetry vector 
$\xi$  onto the directions tangent to the fibres $X_\pm$ over the two poles and, similarly
$J_\pm\equiv J|_{X_\pm}$  the transverse K\"ahler class of the GK geometry restricted to the fibres at the poles then we have 
$\mathcal{V}_{2r+1}^\pm  =    \mathcal{V}_{2r+1} ( \xi_\pm ;  [J_\pm])$. By carefully taking the orientations of 
$\mathcal{V}_{2r+1}^\pm$ into account, both the ``twist" and the ``anti-twist" ways of preserving supersymmetry with spindles
\cite{Ferrero:2021etw} can be realised.
The expression (\ref{ssusytomaster}) provides a
``gravitational block" decomposition of $S_{\text{SUSY}}$.
In addition, results for $\Sigma=S^2$ can also be obtained by carefully taking limits of the spindle case and 
this leads to a gravitational block decomposition of $S_{\text{SUSY}}$ for this case too.

The notion of gravitational blocks was 
proposed in \cite{Hosseini:2019iad} in the context of 
certain supersymmetric black holes in AdS$_4\times S^7$ and black strings in AdS$_5\times S^5$, with spherical horizons,
and then extended to some classes of solutions involving spindles in \cite{Hosseini:2021fge,Faedo:2021nub}.
In these examples it was shown that the entropy can be obtained by extremizing certain entropy functions that are in turn obtained by summing 
basic building blocks, an observation which was inspired by the factorization of 
 partition functions of ${\cal N}=2$ field theories in $d=3$ \cite{Beem:2012mb}.
The results of \cite{Boido:2022mbe}, including \eqref{ssusytomaster}, provide a systematic derivation of the spindle
results of \cite{Hosseini:2021fge,Faedo:2021nub}, as well as extending them 
to the whole class of GK geometry consisting of fibrations of arbitrary SE manifolds over spindles, and, 
furthermore, over two-spheres.

Finally, for GK geometry with $X_{2r+1}$ fibred over a spindle, when $X_{2r+1}$ is toric one can write the supersymmetric action
\eqref{ssusytomaster} in terms of 
toric data and be much 
more explicit about the resulting 
extremization problem.
This was discussed in detail for 
$AdS_3\times Y_7$ and $AdS_2\times Y_9$ solutions in \cite{Boido:2022iye,Boido:2022mbe}.
For the $AdS_3\times Y_7$ examples, it was again proven that there is 
off-shell agreement between the 
geometric and the field theory versions of $c$-extremization, providing another
general exact result for AdS/CFT. 
For some examples in the $AdS_2\times Y_9$ class, the (off-shell) entropy agrees with the field theory computation
of the spindle index \cite{Inglese:2023wky} in the large $N$ limit, thus providing 
a microstate counting interpretation for the entropy of 
accelerating black holes \cite{Colombo:2024mts}.

\section{Final comments}

We have discussed various aspects of SE and GK geometry, highlighting their
similarities and interconnections, as well as the history, without trying to be comprehensive. 
In both cases, physical considerations lead to the formulation of novel extremization problems which have been
instrumental in elucidating the dual SCFTs and deepening our understanding of the geometry. Strikingly,
these extremization principles enable one to compute quantities of physical interest without having a full solution, just inputting some topological conditions and assuming that the solution exists. 

Similar physical considerations indicate that other kinds of geometries of string/M-theory of the form $AdS\times M$ with an R-symmetry should also
admit analogous extremization principles. It has recently been shown in  \cite{BenettiGenolini:2023kxp}
that this is indeed the case, using the ingredients arising from the G-structure programme outlined in the introduction.
In fact, in settings more general than the context of AdS solutions, it has been shown that supersymmetric 
solutions to supergravity theories with an R-symmetry have a set of canonically defined equivariantly closed forms which
can be constructed from the Killing spinor bilinears and 
supergravity fields. 
Furthermore, integrals of these forms can be evaluated using
``localization" i.e. using the Berline-Vergne-Atiyah-Bott 
fixed point formula \cite{BV:1982,Atiyah:1984px} 
which expresses the integrals in terms of the fixed point set of the R-symmetry Killing vector. The precise details of these constructions, as well as the physical observables one can compute, depend on which particular supergravity theory one is studying and which class of solutions one is considering. A number of different examples have now been explored from this point of view
\cite{BenettiGenolini:2023kxp,BenettiGenolini:2023yfe,BenettiGenolini:2023ndb,BenettiGenolini:2024kyy,Couzens:2024vbn,Hristov:2024cgj,BenettiGenolini:2024xeo,Cassani:2024kjn,BenettiGenolini:2024hyd,Crisafio:2024fyc,BenettiGenolini:2024lbj}
and in each case the formalism allows one to profitably analyze general classes of solutions
most of which are unlikely to ever be found in explicit  form. Equivariant localization has also been used to study supergravity solutions from other related points of view in 
\cite{Hristov:2021qsw,Martelli:2023oqk,Colombo:2023fhu,Colombo:2025ihp}. 
We find it remarkable that such general and simple features of supersymmetric solutions of supergravity theories are still being
discovered more than 20 years after the G-structure programme was initiated.

\section*{Acknowledgements}

\noindent 
We thank our many collaborators for the work reviewed here, especially
Chris Couzens, Nakwoo Kim, Daniel Waldram, and Shing-Tung Yau.
This work was supported by STFC grants  ST/X000575/1 and
ST/X000761/1. DM acknowledges partial support by the INFN and by a grant Trapezio (2023) of the Fondazione Compagnia di San Paolo. 
JPG is a Visiting Fellow at the Perimeter Institute. 


\begin{thebibliography}{100}

\bibitem{Maldacena:1997re}
J.~M. Maldacena, ``{The Large N limit of superconformal field theories and
  supergravity},'' \href{http://dx.doi.org/10.4310/ATMP.1998.v2.n2.a1}{{\em
  Adv. Theor. Math. Phys.} {\bfseries 2} (1998) 231--252},
  \href{http://arxiv.org/abs/hep-th/9711200}{{\ttfamily arXiv:hep-th/9711200}}.

\bibitem{Aharony:2008ug}
O.~Aharony, O.~Bergman, D.~L. Jafferis, and J.~Maldacena, ``{N=6 superconformal
  Chern-Simons-matter theories, M2-branes and their gravity duals},''
  \href{http://dx.doi.org/10.1088/1126-6708/2008/10/091}{{\em JHEP} {\bfseries
  10} (2008) 091},
\href{http://arxiv.org/abs/0806.1218}{{\ttfamily arXiv:0806.1218 [hep-th]}}.

\bibitem{Klebanov:1998hh}
I.~R. Klebanov and E.~Witten, ``{Superconformal field theory on three-branes at
  a Calabi-Yau singularity},''
  \href{http://dx.doi.org/10.1016/S0550-3213(98)00654-3}{{\em Nucl.Phys.}
  {\bfseries B536} (1998) 199--218},
\href{http://arxiv.org/abs/hep-th/9807080}{{\ttfamily arXiv:hep-th/9807080
  [hep-th]}}.

\bibitem{Acharya:1998db}
B.~S. Acharya, J.~M. Figueroa-O'Farrill, C.~M. Hull, and B.~J. Spence,
  ``{Branes at conical singularities and holography},''
  \href{http://dx.doi.org/10.4310/ATMP.1998.v2.n6.a2}{{\em Adv. Theor. Math.
  Phys.} {\bfseries 2} (1999) 1249--1286},
\href{http://arxiv.org/abs/hep-th/9808014}{{\ttfamily arXiv:hep-th/9808014
  [hep-th]}}.

\bibitem{Morrison:1998cs}
D.~R. Morrison and M.~R. Plesser, ``{Nonspherical horizons. 1.},''
  \href{http://dx.doi.org/10.4310/ATMP.1999.v3.n1.a1}{{\em Adv. Theor. Math.
  Phys.} {\bfseries 3} (1999) 1--81},
  \href{http://arxiv.org/abs/hep-th/9810201}{{\ttfamily arXiv:hep-th/9810201}}.

\bibitem{Freund:1980xh}
P.~G.~O. Freund and M.~A. Rubin, ``{Dynamics of Dimensional Reduction},''
  \href{http://dx.doi.org/10.1016/0370-2693(80)90590-0}{{\em Phys. Lett. B}
  {\bfseries 97} (1980) 233--235}.

\bibitem{Castellani:1983yg}
L.~Castellani, L.~J. Romans, and N.~P. Warner, ``{A Classification of
  Compactifying Solutions for $d=11$ Supergravity},''
  \href{http://dx.doi.org/10.1016/0550-3213(84)90055-5}{{\em Nucl. Phys. B}
  {\bfseries 241} (1984) 429--462}.

\bibitem{Romans:1984an}
L.~Romans, ``{New compactifications of chiral N=2 d=10 supergravity},''
\href{http://dx.doi.org/10.1016/0370-2693(85)90479-4}{{\em Phys.Lett.}
  {\bfseries B153} (1985) 392}.

\bibitem{Gauntlett:2004yd}
J.~P. Gauntlett, D.~Martelli, J.~Sparks, and D.~Waldram, ``{Sasaki-Einstein
  metrics on $S^2 \times S^3$},''
  \href{http://dx.doi.org/10.4310/ATMP.2004.v8.n4.a3}{{\em Adv. Theor. Math.
  Phys.} {\bfseries 8} no.~4, (2004) 711--734},
\href{http://arxiv.org/abs/hep-th/0403002}{{\ttfamily arXiv:hep-th/0403002
  [hep-th]}}.

\bibitem{CheegerJeff1994Otcs}
J.~Cheeger and G.~Tian, ``On the cone structure at infinity of ricci flat
  manifolds with euclidean volume growth and quadratic curvature decay,'' {\em
  Inventiones mathematicae} {\bfseries 118} no.~1, (1994) 493--571.

\bibitem{Benvenuti:2004dy}
S.~Benvenuti, S.~Franco, A.~Hanany, D.~Martelli, and J.~Sparks, ``{An Infinite
  family of superconformal quiver gauge theories with Sasaki-Einstein duals},''
  \href{http://dx.doi.org/10.1088/1126-6708/2005/06/064}{{\em JHEP} {\bfseries
  06} (2005) 064},
\href{http://arxiv.org/abs/hep-th/0411264}{{\ttfamily arXiv:hep-th/0411264
  [hep-th]}}.

\bibitem{Intriligator:2003jj}
K.~A. Intriligator and B.~Wecht, ``{The Exact superconformal R symmetry
  maximizes $a$},'' \href{http://dx.doi.org/10.1016/S0550-3213(03)00459-0}{{\em
  Nucl. Phys.} {\bfseries B667} (2003) 183--200},
\href{http://arxiv.org/abs/hep-th/0304128}{{\ttfamily arXiv:hep-th/0304128
  [hep-th]}}.

\bibitem{Martelli:2005tp}
D.~Martelli, J.~Sparks, and S.-T. Yau, ``{The Geometric dual of
  $a$-maximisation for Toric Sasaki-Einstein manifolds},''
  \href{http://dx.doi.org/10.1007/s00220-006-0087-0}{{\em Commun. Math. Phys.}
  {\bfseries 268} (2006) 39--65},
\href{http://arxiv.org/abs/hep-th/0503183}{{\ttfamily arXiv:hep-th/0503183
  [hep-th]}}.

\bibitem{Martelli:2006yb}
D.~Martelli, J.~Sparks, and S.-T. Yau, ``{Sasaki-Einstein manifolds and volume
  minimisation},'' \href{http://dx.doi.org/10.1007/s00220-008-0479-4}{{\em
  Commun. Math. Phys.} {\bfseries 280} (2008) 611--673},
\href{http://arxiv.org/abs/hep-th/0603021}{{\ttfamily arXiv:hep-th/0603021
  [hep-th]}}.

\bibitem{Jafferis:2010un}
D.~L. Jafferis, ``{The Exact Superconformal R-Symmetry Extremizes Z},''
  \href{http://dx.doi.org/10.1007/JHEP05(2012)159}{{\em JHEP} {\bfseries 05}
  (2012) 159},
\href{http://arxiv.org/abs/1012.3210}{{\ttfamily arXiv:1012.3210 [hep-th]}}.

\bibitem{Gauntlett:2001ur}
J.~P. Gauntlett, N.~Kim, D.~Martelli, and D.~Waldram, ``{Five-branes wrapped on
  SLAG three cycles and related geometry},''
  \href{http://dx.doi.org/10.1088/1126-6708/2001/11/018}{{\em JHEP} {\bfseries
  11} (2001) 018}, \href{http://arxiv.org/abs/hep-th/0110034}{{\ttfamily
  arXiv:hep-th/0110034}}.

\bibitem{Gauntlett:2002sc}
J.~P. Gauntlett, D.~Martelli, S.~Pakis, and D.~Waldram, ``{G structures and
  wrapped NS5-branes},''
  \href{http://dx.doi.org/10.1007/s00220-004-1066-y}{{\em Commun. Math. Phys.}
  {\bfseries 247} (2004) 421--445},
  \href{http://arxiv.org/abs/hep-th/0205050}{{\ttfamily arXiv:hep-th/0205050}}.

\bibitem{Gauntlett:2003cy}
J.~P. Gauntlett, D.~Martelli, and D.~Waldram, ``{Superstrings with intrinsic
  torsion},'' \href{http://dx.doi.org/10.1103/PhysRevD.69.086002}{{\em Phys.
  Rev.} {\bfseries D69} (2004) 086002},
\href{http://arxiv.org/abs/hep-th/0302158}{{\ttfamily arXiv:hep-th/0302158
  [hep-th]}}.

\bibitem{Gauntlett:2002fz}
J.~P. Gauntlett and S.~Pakis, ``{The Geometry of D = 11 killing spinors},''
  \href{http://dx.doi.org/10.1088/1126-6708/2003/04/039}{{\em JHEP} {\bfseries
  04} (2003) 039}, \href{http://arxiv.org/abs/hep-th/0212008}{{\ttfamily
  arXiv:hep-th/0212008}}.

\bibitem{Gauntlett:2003wb}
J.~P. Gauntlett, J.~B. Gutowski, and S.~Pakis, ``{The Geometry of D = 11 null
  Killing spinors},''
  \href{http://dx.doi.org/10.1088/1126-6708/2003/12/049}{{\em JHEP} {\bfseries
  12} (2003) 049}, \href{http://arxiv.org/abs/hep-th/0311112}{{\ttfamily
  arXiv:hep-th/0311112}}.

\bibitem{Gauntlett:2002nw}
J.~P. Gauntlett, J.~B. Gutowski, C.~M. Hull, S.~Pakis, and H.~S. Reall, ``{All
  supersymmetric solutions of minimal supergravity in five- dimensions},''
  \href{http://dx.doi.org/10.1088/0264-9381/20/21/005}{{\em Class. Quant.
  Grav.} {\bfseries 20} (2003) 4587--4634},
  \href{http://arxiv.org/abs/hep-th/0209114}{{\ttfamily arXiv:hep-th/0209114}}.

\bibitem{Gauntlett:2003fk}
J.~P. Gauntlett and J.~B. Gutowski, ``{All supersymmetric solutions of minimal
  gauged supergravity in five-dimensions},''
  \href{http://dx.doi.org/10.1103/PhysRevD.70.089901,
  10.1103/PhysRevD.68.105009}{{\em Phys. Rev.} {\bfseries D68} (2003) 105009},
  \href{http://arxiv.org/abs/hep-th/0304064}{{\ttfamily arXiv:hep-th/0304064
  [hep-th]}}.
[Erratum: Phys. Rev.D70,089901(2004)].

\bibitem{Elvang:2004rt}
H.~Elvang, R.~Emparan, D.~Mateos, and H.~S. Reall, ``{A Supersymmetric black
  ring},'' \href{http://dx.doi.org/10.1103/PhysRevLett.93.211302}{{\em Phys.
  Rev. Lett.} {\bfseries 93} (2004) 211302},
  \href{http://arxiv.org/abs/hep-th/0407065}{{\ttfamily arXiv:hep-th/0407065}}.

\bibitem{Gauntlett:2004qy}
J.~P. Gauntlett and J.~B. Gutowski, ``{General concentric black rings},''
  \href{http://dx.doi.org/10.1103/PhysRevD.71.045002}{{\em Phys. Rev. D}
  {\bfseries 71} (2005) 045002},
  \href{http://arxiv.org/abs/hep-th/0408122}{{\ttfamily arXiv:hep-th/0408122}}.

\bibitem{Bena:2007kg}
I.~Bena and N.~P. Warner, ``{Black holes, black rings and their microstates},''
  \href{http://dx.doi.org/10.1007/978-3-540-79523-0_1}{{\em Lect. Notes Phys.}
  {\bfseries 755} (2008) 1--92},
  \href{http://arxiv.org/abs/hep-th/0701216}{{\ttfamily arXiv:hep-th/0701216}}.

\bibitem{Gauntlett:2004zh}
J.~P. Gauntlett, D.~Martelli, J.~Sparks, and D.~Waldram, ``{Supersymmetric
  AdS(5) solutions of M-theory},''
  \href{http://dx.doi.org/10.1088/0264-9381/21/18/005}{{\em Class. Quant.
  Grav.} {\bfseries 21} (2004) 4335--4366},
\href{http://arxiv.org/abs/hep-th/0402153}{{\ttfamily arXiv:hep-th/0402153}}.

\bibitem{Kim:2005ez}
N.~Kim, ``{AdS$_3$ solutions of IIB supergravity from D3-branes},''
  \href{http://dx.doi.org/10.1088/1126-6708/2006/01/094}{{\em JHEP} {\bfseries
  01} (2006) 094},
\href{http://arxiv.org/abs/hep-th/0511029}{{\ttfamily arXiv:hep-th/0511029
  [hep-th]}}.

\bibitem{Kim:2006qu}
N.~Kim and J.-D. Park, ``{Comments on AdS(2) solutions of D = 11
  supergravity},'' \href{http://dx.doi.org/10.1088/1126-6708/2006/09/041}{{\em
  JHEP} {\bfseries 09} (2006) 041},
\href{http://arxiv.org/abs/hep-th/0607093}{{\ttfamily arXiv:hep-th/0607093}}.

\bibitem{Gauntlett:2007ts}
J.~P. Gauntlett and N.~Kim, ``{Geometries with Killing Spinors and
  Supersymmetric AdS Solutions},''
  \href{http://dx.doi.org/10.1007/s00220-008-0575-5}{{\em Commun. Math. Phys.}
  {\bfseries 284} (2008) 897--918},
\href{http://arxiv.org/abs/0710.2590}{{\ttfamily arXiv:0710.2590 [hep-th]}}.

\bibitem{Couzens:2018wnk}
C.~Couzens, J.~P. Gauntlett, D.~Martelli, and J.~Sparks, ``{A geometric dual of
  $c$-extremization},'' \href{http://dx.doi.org/10.1007/JHEP01(2019)212}{{\em
  JHEP} {\bfseries 01} (2019) 212},
\href{http://arxiv.org/abs/1810.11026}{{\ttfamily arXiv:1810.11026 [hep-th]}}.

\bibitem{Gauntlett:2018dpc}
J.~P. Gauntlett, D.~Martelli, and J.~Sparks, ``{Toric geometry and the dual of
  $c$-extremization},'' \href{http://dx.doi.org/10.1007/JHEP01(2019)204}{{\em
  JHEP} {\bfseries 01} (2019) 204},
\href{http://arxiv.org/abs/1812.05597}{{\ttfamily arXiv:1812.05597 [hep-th]}}.

\bibitem{Hosseini:2019use}
S.~M. Hosseini and A.~Zaffaroni, ``{Proving the equivalence of
  $c$-extremization and its gravitational dual for all toric quivers},''
  \href{http://dx.doi.org/10.1007/JHEP03(2019)108}{{\em JHEP} {\bfseries 03}
  (2019) 108},
\href{http://arxiv.org/abs/1901.05977}{{\ttfamily arXiv:1901.05977 [hep-th]}}.

\bibitem{Hosseini:2019ddy}
S.~M. Hosseini and A.~Zaffaroni, ``{Geometry of $\mathcal{I}$-extremization and
  black holes microstates},''
  \href{http://dx.doi.org/10.1007/JHEP07(2019)174}{{\em JHEP} {\bfseries 07}
  (2019) 174},
\href{http://arxiv.org/abs/1904.04269}{{\ttfamily arXiv:1904.04269 [hep-th]}}.

\bibitem{Gauntlett:2019roi}
J.~P. Gauntlett, D.~Martelli, and J.~Sparks, ``{Toric geometry and the dual of
  ${\cal I}$-extremization},''
  \href{http://dx.doi.org/10.1007/JHEP06(2019)140}{{\em JHEP} {\bfseries 06}
  (2019) 140},
\href{http://arxiv.org/abs/1904.04282}{{\ttfamily arXiv:1904.04282 [hep-th]}}.

\bibitem{Kim:2019umc}
H.~Kim and N.~Kim, ``{Black holes with baryonic charge and
  $\mathcal{I}$-extremization},''
  \href{http://dx.doi.org/10.1007/JHEP11(2019)050}{{\em JHEP} {\bfseries 11}
  (2019) 050}, \href{http://arxiv.org/abs/1904.05344}{{\ttfamily
  arXiv:1904.05344 [hep-th]}}.

\bibitem{Benini:2012cz}
F.~Benini and N.~Bobev, ``{Exact two-dimensional superconformal R-symmetry and
  c-extremization},''
  \href{http://dx.doi.org/10.1103/PhysRevLett.110.061601}{{\em Phys. Rev.
  Lett.} {\bfseries 110} no.~6, (2013) 061601},
\href{http://arxiv.org/abs/1211.4030}{{\ttfamily arXiv:1211.4030 [hep-th]}}.

\bibitem{Benini:2013cda}
F.~Benini and N.~Bobev, ``{Two-dimensional SCFTs from wrapped branes and
  c-extremization},'' \href{http://dx.doi.org/10.1007/JHEP06(2013)005}{{\em
  JHEP} {\bfseries 06} (2013) 005},
\href{http://arxiv.org/abs/1302.4451}{{\ttfamily arXiv:1302.4451 [hep-th]}}.

\bibitem{Benini:2015eyy}
F.~Benini, K.~Hristov, and A.~Zaffaroni, ``{Black hole microstates in AdS$_{4}$
  from supersymmetric localization},''
  \href{http://dx.doi.org/10.1007/JHEP05(2016)054}{{\em JHEP} {\bfseries 05}
  (2016) 054},
\href{http://arxiv.org/abs/1511.04085}{{\ttfamily arXiv:1511.04085 [hep-th]}}.

\bibitem{Boyer:2008era}
C.~P. Boyer and K.~Galicki, {\em {Sasakian Geometry}}.
\newblock Oxford University Press, Oxford,
2008.
\newblock

\bibitem{Sparks:2010sn}
J.~Sparks, ``{Sasaki-Einstein Manifolds},''
  \href{http://dx.doi.org/10.4310/SDG.2011.v16.n1.a6}{{\em Surveys Diff. Geom.}
  {\bfseries 16} (2011) 265--324},
\href{http://arxiv.org/abs/1004.2461}{{\ttfamily arXiv:1004.2461 [math.DG]}}.

\bibitem{FRIEDRICHT1989Emod}
T.~Friedrich and I.~Kath, ``Einstein manifolds of dimension five with small
  first eigenvalue of the dirac operator,''
  \href{http://dx.doi.org/10.4310/jdg/1214442874}{{\em Journal of differential
  geometry} {\bfseries 29} no.~2, (1989) 263--279}.

\bibitem{Conti:2006dg}
D.~Conti, ``{Cohomogeneity one Einstein-Sasaki 5-manifolds},''
  \href{http://dx.doi.org/10.1007/s00220-007-0286-3}{{\em Commun. Math. Phys.}
  {\bfseries 274} (2007) 751--774},
  \href{http://arxiv.org/abs/math/0606323}{{\ttfamily arXiv:math/0606323}}.

\bibitem{bergery}
L.~B\'erard-Bergery {\em Institut Elie Cartan} {\bfseries 6} (1982) 1.

\bibitem{Page:1985bq}
D.~N. Page and C.~N. Pope, ``{Inhomogeneous Einstein metrics in complex line
  bundles},'' \href{http://dx.doi.org/10.1088/0264-9381/4/2/005}{{\em Class.
  Quant. Grav.} {\bfseries 4} (1987) 213--225}.

\bibitem{Gauntlett:2004hh}
J.~P. Gauntlett, D.~Martelli, J.~F. Sparks, and D.~Waldram, ``{A New infinite
  class of Sasaki-Einstein manifolds},''
  \href{http://dx.doi.org/10.4310/ATMP.2004.v8.n6.a3}{{\em Adv. Theor. Math.
  Phys.} {\bfseries 8} no.~6, (2004) 987--1000},
\href{http://arxiv.org/abs/hep-th/0403038}{{\ttfamily arXiv:hep-th/0403038
  [hep-th]}}.

\bibitem{Cvetic:2005ft}
M.~Cvetic, H.~Lu, D.~N. Page, and C.~N. Pope, ``{New Einstein-Sasaki spaces in
  five and higher dimensions},''
  \href{http://dx.doi.org/10.1103/PhysRevLett.95.071101}{{\em Phys. Rev. Lett.}
  {\bfseries 95} (2005) 071101},
\href{http://arxiv.org/abs/hep-th/0504225}{{\ttfamily arXiv:hep-th/0504225
  [hep-th]}}.

\bibitem{Houri:2007xz}
T.~Houri, T.~Oota, and Y.~Yasui, ``{Closed conformal Killing-Yano tensor and
  Kerr-NUT-de Sitter spacetime uniqueness},''
  \href{http://dx.doi.org/10.1016/j.physletb.2007.09.034}{{\em Phys. Lett. B}
  {\bfseries 656} (2007) 214--216},
  \href{http://arxiv.org/abs/0708.1368}{{\ttfamily arXiv:0708.1368 [hep-th]}}.

\bibitem{Hawking:1998kw}
S.~W. Hawking, C.~J. Hunter, and M.~Taylor, ``{Rotation and the AdS / CFT
  correspondence},'' \href{http://dx.doi.org/10.1103/PhysRevD.59.064005}{{\em
  Phys. Rev. D} {\bfseries 59} (1999) 064005},
  \href{http://arxiv.org/abs/hep-th/9811056}{{\ttfamily arXiv:hep-th/9811056}}.

\bibitem{Martelli:2005wy}
D.~Martelli and J.~Sparks, ``{Toric Sasaki-Einstein metrics on $S^2 \times
  S^3$},'' \href{http://dx.doi.org/10.1016/j.physletb.2005.06.059}{{\em Phys.
  Lett.} {\bfseries B621} (2005) 208--212},
\href{http://arxiv.org/abs/hep-th/0505027}{{\ttfamily arXiv:hep-th/0505027
  [hep-th]}}.

\bibitem{Guillemin1994KaehlerSO}
V.~W. Guillemin, ``Kaehler structures on toric varieties,'' {\em Journal of
  Differential Geometry} {\bfseries 40} (1994) 285--309.

\bibitem{abreu2000kahler}
M.~Abreu, ``Kahler geometry of toric manifolds in symplectic coordinates,''
  {\em Fields Inst. Commun.} {\bfseries 35} (08, 2000) ,
  \href{http://arxiv.org/abs/{math/0004122}}{{\ttfamily {arXiv}:{math/0004122}
  [{math.DG}]}}.

\bibitem{Abreu:2001to}
M.~Abreu, ``Kahler metrics on toric orbifolds,'' {\em Journal of Differential
  Geometry} {\bfseries 58} (2001) 151--187,
  \href{http://arxiv.org/abs/math/0105112}{{\ttfamily arXiv:math/0105112
  [math.DG]}}.

\bibitem{Burns2005KhlerMO}
D.~Burns, V.~W. Guillemin, and E.~Lerman, ``K{\"a}hler metrics on singular
  toric varieties,'' {\em Pacific Journal of Mathematics} {\bfseries 238}
  (2005) 27--40. \url{https://api.semanticscholar.org/CorpusID:6811372}.

\bibitem{Futaki:2006cc}
A.~Futaki, H.~Ono, and G.~Wang, ``{Transverse Kahler geometry of Sasaki
  manifolds and toric Sasaki-Einstein manifolds},'' {\em J. Diff. Geom.}
  {\bfseries 83} no.~3, (2009) 585--636,
  \href{http://arxiv.org/abs/math/0607586}{{\ttfamily arXiv:math/0607586}}.

\bibitem{Gauntlett:2019pqg}
J.~P. Gauntlett, D.~Martelli, and J.~Sparks, ``{Fibred GK geometry and
  supersymmetric $AdS$ solutions},''
  \href{http://dx.doi.org/10.1007/JHEP11(2019)176}{{\em JHEP} {\bfseries 11}
  (2019) 176}, \href{http://arxiv.org/abs/1910.08078}{{\ttfamily
  arXiv:1910.08078 [hep-th]}}.

\bibitem{Gauntlett:2006vf}
J.~P. Gauntlett, D.~Martelli, J.~Sparks, and S.-T. Yau, ``{Obstructions to the
  existence of Sasaki-Einstein metrics},''
  \href{http://dx.doi.org/10.1007/s00220-007-0213-7}{{\em Commun. Math. Phys.}
  {\bfseries 273} (2007) 803--827},
\href{http://arxiv.org/abs/hep-th/0607080}{{\ttfamily arXiv:hep-th/0607080
  [hep-th]}}.

\bibitem{Komargodski:2011vj}
Z.~Komargodski and A.~Schwimmer, ``{On Renormalization Group Flows in Four
  Dimensions},'' \href{http://dx.doi.org/10.1007/JHEP12(2011)099}{{\em JHEP}
  {\bfseries 12} (2011) 099}, \href{http://arxiv.org/abs/1107.3987}{{\ttfamily
  arXiv:1107.3987 [hep-th]}}.

\bibitem{Ross:2011big}
J.~Ross and R.~Thomas, ``{Weighted projective embeddings, stability of
  orbifolds and constant scalar curvature K\"ahler metrics},'' {\em J. Diff.
  Geom.} {\bfseries 88} no.~1, (2011) 109--159,
  \href{http://arxiv.org/abs/0907.5214}{{\ttfamily arXiv:0907.5214 [math.AG]}}.

\bibitem{Collins:2012dh}
T.~C. Collins and G.~Szekelyhidi, ``{K-Semistability for irregular Sasakian
  manifolds},'' \href{http://dx.doi.org/10.4310/jdg/1525399217}{{\em J. Diff.
  Geom.} {\bfseries 109} no.~1, (2018) 81--109},
  \href{http://arxiv.org/abs/1204.2230}{{\ttfamily arXiv:1204.2230 [math.DG]}}.

\bibitem{Collins:2015qsb}
T.~C. Collins and G.~Sz\'ekelyhidi, ``{Sasaki-Einstein metrics and
  K-stability},'' {\em Geom. and Topol.} no.~23, (2019) 1339--1413,
\href{http://arxiv.org/abs/1512.07213}{{\ttfamily arXiv:1512.07213 [math.DG]}}.

\bibitem{Collins:2016icw}
T.~C. Collins, D.~Xie, and S.-T. Yau, ``{K stability and stability of chiral
  ring},'' \href{http://arxiv.org/abs/1606.09260}{{\ttfamily arXiv:1606.09260
  [hep-th]}}.

\bibitem{Douglas:1996sw}
M.~R. Douglas and G.~W. Moore, ``{D-branes, quivers, and ALE instantons},''
  \href{http://arxiv.org/abs/hep-th/9603167}{{\ttfamily arXiv:hep-th/9603167}}.

\bibitem{Feng:2000mi}
B.~Feng, A.~Hanany, and Y.-H. He, ``{D-brane gauge theories from toric
  singularities and toric duality},''
  \href{http://dx.doi.org/10.1016/S0550-3213(00)00699-4}{{\em Nucl. Phys. B}
  {\bfseries 595} (2001) 165--200},
  \href{http://arxiv.org/abs/hep-th/0003085}{{\ttfamily arXiv:hep-th/0003085}}.

\bibitem{Wijnholt:2002qz}
M.~Wijnholt, ``{Large volume perspective on branes at singularities},''
  \href{http://dx.doi.org/10.4310/ATMP.2003.v7.n6.a6}{{\em Adv. Theor. Math.
  Phys.} {\bfseries 7} no.~6, (2003) 1117--1153},
  \href{http://arxiv.org/abs/hep-th/0212021}{{\ttfamily arXiv:hep-th/0212021}}.

\bibitem{Martelli:2004wu}
D.~Martelli and J.~Sparks, ``{Toric geometry, Sasaki-Einstein manifolds and a
  new infinite class of AdS/CFT duals},''
  \href{http://dx.doi.org/10.1007/s00220-005-1425-3}{{\em Commun. Math. Phys.}
  {\bfseries 262} (2006) 51--89},
\href{http://arxiv.org/abs/hep-th/0411238}{{\ttfamily arXiv:hep-th/0411238
  [hep-th]}}.

\bibitem{Bertolini:2004xf}
M.~Bertolini, F.~Bigazzi, and A.~L. Cotrone, ``{New checks and subtleties for
  AdS/CFT and a-maximization},''
  \href{http://dx.doi.org/10.1088/1126-6708/2004/12/024}{{\em JHEP} {\bfseries
  12} (2004) 024}, \href{http://arxiv.org/abs/hep-th/0411249}{{\ttfamily
  arXiv:hep-th/0411249}}.

\bibitem{Herzog:2006bu}
C.~P. Herzog and R.~L. Karp, ``{On the geometry of quiver gauge theories
  (Stacking exceptional collections)},''
  \href{http://dx.doi.org/10.4310/ATMP.2009.v13.n3.a1}{{\em Adv. Theor. Math.
  Phys.} {\bfseries 13} no.~3, (2009) 599--636},
  \href{http://arxiv.org/abs/hep-th/0605177}{{\ttfamily arXiv:hep-th/0605177}}.

\bibitem{Franco:2005sm}
S.~Franco, A.~Hanany, D.~Martelli, J.~Sparks, D.~Vegh, and B.~Wecht, ``{Gauge
  theories from toric geometry and brane tilings},''
  \href{http://dx.doi.org/10.1088/1126-6708/2006/01/128}{{\em JHEP} {\bfseries
  01} (2006) 128},
\href{http://arxiv.org/abs/hep-th/0505211}{{\ttfamily arXiv:hep-th/0505211
  [hep-th]}}.

\bibitem{Franco:2005rj}
S.~Franco, A.~Hanany, K.~D. Kennaway, D.~Vegh, and B.~Wecht, ``{Brane dimers
  and quiver gauge theories},''
  \href{http://dx.doi.org/10.1088/1126-6708/2006/01/096}{{\em JHEP} {\bfseries
  01} (2006) 096}, \href{http://arxiv.org/abs/hep-th/0504110}{{\ttfamily
  arXiv:hep-th/0504110}}.

\bibitem{Hanany:2008fj}
A.~Hanany, D.~Vegh, and A.~Zaffaroni, ``{Brane Tilings and M2 Branes},''
  \href{http://dx.doi.org/10.1088/1126-6708/2009/03/012}{{\em JHEP} {\bfseries
  03} (2009) 012},
\href{http://arxiv.org/abs/0809.1440}{{\ttfamily arXiv:0809.1440 [hep-th]}}.

\bibitem{Franco:2017jeo}
S.~Franco, Y.-H. He, C.~Sun, and Y.~Xiao, ``{A Comprehensive Survey of Brane
  Tilings},'' \href{http://dx.doi.org/10.1142/S0217751X17501421}{{\em Int. J.
  Mod. Phys.} {\bfseries A32} no.~23n24, (2017) 1750142},
\href{http://arxiv.org/abs/1702.03958}{{\ttfamily arXiv:1702.03958 [hep-th]}}.

\bibitem{Butti:2005vn}
A.~Butti and A.~Zaffaroni, ``{R-charges from toric diagrams and the equivalence
  of a-maximization and Z-minimization},''
  \href{http://dx.doi.org/10.1088/1126-6708/2005/11/019}{{\em JHEP} {\bfseries
  11} (2005) 019},
\href{http://arxiv.org/abs/hep-th/0506232}{{\ttfamily arXiv:hep-th/0506232
  [hep-th]}}.

\bibitem{Eager:2010yu}
R.~Eager, ``{Equivalence of A-Maximization and Volume Minimization},''
  \href{http://dx.doi.org/10.1007/JHEP01(2014)089}{{\em JHEP} {\bfseries 01}
  (2014) 089}, \href{http://arxiv.org/abs/1011.1809}{{\ttfamily arXiv:1011.1809
  [hep-th]}}.

\bibitem{Eager:2012hx}
R.~Eager, J.~Schmude, and Y.~Tachikawa, ``{Superconformal Indices,
  Sasaki-Einstein Manifolds, and Cyclic Homologies},''
  \href{http://dx.doi.org/10.4310/ATMP.2014.v18.n1.a3}{{\em Adv. Theor. Math.
  Phys.} {\bfseries 18} no.~1, (2014) 129--175},
  \href{http://arxiv.org/abs/1207.0573}{{\ttfamily arXiv:1207.0573 [hep-th]}}.

\bibitem{Boyer:2000pg}
C.~P. Boyer, K.~Galicki, and M.~Nakamaye, ``{On the geometry of
  Sasakian-Einstein 5 manifolds},''
  \href{http://dx.doi.org/10.1007/s00208-002-0388-3}{{\em Math. Ann.}
  {\bfseries 325} (2003) 485--524},
  \href{http://arxiv.org/abs/math/0012047}{{\ttfamily arXiv:math/0012047}}.

\bibitem{BoyerCharlesP.2001NEMi}
C.~P. Boyer and K.~Galicki, ``New einstein metrics in dimension five,'' {\em
  Journal of differential geometry} {\bfseries 57} no.~3, (2001) 443--463.

\bibitem{suss2021}
H.~S\"u\ss, ``On irregular sasaki-einstein metrics in dimension 5,''
  \href{http://dx.doi.org/10.1070/SM9487}{{\em Sbornik: Mathematics} {\bfseries
  212} no.~9, (Sep, 2021) 1261}.

\bibitem{Fazzi:2019gvt}
M.~Fazzi and A.~Tomasiello, ``{Holography, Matrix Factorizations and
  K-stability},'' \href{http://dx.doi.org/10.1007/JHEP05(2020)119}{{\em JHEP}
  {\bfseries 05} (2020) 119}, \href{http://arxiv.org/abs/1906.08272}{{\ttfamily
  arXiv:1906.08272 [hep-th]}}.

\bibitem{Fabbri:1999hw}
D.~Fabbri, P.~Fre', L.~Gualtieri, C.~Reina, A.~Tomasiello, {\em et al.}, ``{3-D
  superconformal theories from Sasakian seven manifolds: New nontrivial
  evidences for AdS(4) / CFT(3)},''
  \href{http://dx.doi.org/10.1016/S0550-3213(00)00098-5}{{\em Nucl.Phys.}
  {\bfseries B577} (2000) 547--608},
\href{http://arxiv.org/abs/hep-th/9907219}{{\ttfamily arXiv:hep-th/9907219
  [hep-th]}}.

\bibitem{Ceresole:1999zg}
A.~Ceresole, G.~Dall'Agata, R.~D'Auria, and S.~Ferrara, ``{M theory on the
  Stiefel manifold and 3-D conformal field theories},''
  \href{http://dx.doi.org/10.1088/1126-6708/2000/03/011}{{\em JHEP} {\bfseries
  03} (2000) 011}, \href{http://arxiv.org/abs/hep-th/9912107}{{\ttfamily
  arXiv:hep-th/9912107}}.

\bibitem{Schwarz:2004yj}
J.~H. Schwarz, ``{Superconformal Chern-Simons theories},''
  \href{http://dx.doi.org/10.1088/1126-6708/2004/11/078}{{\em JHEP} {\bfseries
  11} (2004) 078}, \href{http://arxiv.org/abs/hep-th/0411077}{{\ttfamily
  arXiv:hep-th/0411077}}.

\bibitem{Bagger:2006sk}
J.~Bagger and N.~Lambert, ``{Modeling Multiple M2's},''
  \href{http://dx.doi.org/10.1103/PhysRevD.75.045020}{{\em Phys. Rev. D}
  {\bfseries 75} (2007) 045020},
  \href{http://arxiv.org/abs/hep-th/0611108}{{\ttfamily arXiv:hep-th/0611108}}.

\bibitem{Hosomichi:2008jd}
K.~Hosomichi, K.-M. Lee, S.~Lee, S.~Lee, and J.~Park, ``{N=4 Superconformal
  Chern-Simons Theories with Hyper and Twisted Hyper Multiplets},''
  \href{http://dx.doi.org/10.1088/1126-6708/2008/07/091}{{\em JHEP} {\bfseries
  07} (2008) 091}, \href{http://arxiv.org/abs/0805.3662}{{\ttfamily
  arXiv:0805.3662 [hep-th]}}.

\bibitem{Martelli:2008si}
D.~Martelli and J.~Sparks, ``{Moduli spaces of Chern-Simons quiver gauge
  theories and AdS(4)/CFT(3)},''
  \href{http://dx.doi.org/10.1103/PhysRevD.78.126005}{{\em Phys. Rev. D}
  {\bfseries 78} (2008) 126005},
  \href{http://arxiv.org/abs/0808.0912}{{\ttfamily arXiv:0808.0912 [hep-th]}}.

\bibitem{Jafferis:2008qz}
D.~L. Jafferis and A.~Tomasiello, ``{A Simple class of N=3 gauge/gravity
  duals},'' \href{http://dx.doi.org/10.1088/1126-6708/2008/10/101}{{\em JHEP}
  {\bfseries 10} (2008) 101}, \href{http://arxiv.org/abs/0808.0864}{{\ttfamily
  arXiv:0808.0864 [hep-th]}}.

\bibitem{Martelli:2011qj}
D.~Martelli and J.~Sparks, ``{The large N limit of quiver matrix models and
  Sasaki-Einstein manifolds},''
  \href{http://dx.doi.org/10.1103/PhysRevD.84.046008}{{\em Phys. Rev.}
  {\bfseries D84} (2011) 046008},
\href{http://arxiv.org/abs/1102.5289}{{\ttfamily arXiv:1102.5289 [hep-th]}}.

\bibitem{Jafferis:2011zi}
D.~L. Jafferis, I.~R. Klebanov, S.~S. Pufu, and B.~R. Safdi, ``{Towards the
  F-Theorem: N=2 Field Theories on the Three-Sphere},''
  \href{http://dx.doi.org/10.1007/JHEP06(2011)102}{{\em JHEP} {\bfseries 06}
  (2011) 102},
\href{http://arxiv.org/abs/1103.1181}{{\ttfamily arXiv:1103.1181 [hep-th]}}.

\bibitem{Benini:2009qs}
F.~Benini, C.~Closset, and S.~Cremonesi, ``{Chiral flavors and M2-branes at
  toric CY4 singularities},''
  \href{http://dx.doi.org/10.1007/JHEP02(2010)036}{{\em JHEP} {\bfseries 02}
  (2010) 036},
\href{http://arxiv.org/abs/0911.4127}{{\ttfamily arXiv:0911.4127 [hep-th]}}.

\bibitem{Benini:2011cma}
F.~Benini, C.~Closset, and S.~Cremonesi, ``{Quantum moduli space of
  Chern-Simons quivers, wrapped D6-branes and AdS4/CFT3},''
  \href{http://dx.doi.org/10.1007/JHEP09(2011)005}{{\em JHEP} {\bfseries 09}
  (2011) 005}, \href{http://arxiv.org/abs/1105.2299}{{\ttfamily arXiv:1105.2299
  [hep-th]}}.

\bibitem{Closset:2012ep}
C.~Closset and S.~Cremonesi, ``{Toric Fano varieties and Chern-Simons
  quivers},'' \href{http://dx.doi.org/10.1007/JHEP05(2012)060}{{\em JHEP}
  {\bfseries 05} (2012) 060}, \href{http://arxiv.org/abs/1201.2431}{{\ttfamily
  arXiv:1201.2431 [hep-th]}}.

\bibitem{Amariti:2012tj}
A.~Amariti and S.~Franco, ``{Free Energy vs Sasaki-Einstein Volume for Infinite
  Families of M2-Brane Theories},''
  \href{http://dx.doi.org/10.1007/JHEP09(2012)034}{{\em JHEP} {\bfseries 09}
  (2012) 034},
\href{http://arxiv.org/abs/1204.6040}{{\ttfamily arXiv:1204.6040 [hep-th]}}.

\bibitem{Martelli:2009ga}
D.~Martelli and J.~Sparks, ``{AdS$_4$/CFT$_3$ duals from M2-branes at
  hypersurface singularities and their deformations},''
  \href{http://dx.doi.org/10.1088/1126-6708/2009/12/017}{{\em JHEP} {\bfseries
  12} (2009) 017},
\href{http://arxiv.org/abs/0909.2036}{{\ttfamily arXiv:0909.2036 [hep-th]}}.

\bibitem{Jafferis:2009th}
D.~L. Jafferis, ``{Quantum corrections to $\mathcal{N} = 2$ Chern-Simons
  theories with flavor and their AdS$_{4}$ duals},''
  \href{http://dx.doi.org/10.1007/JHEP08(2013)046}{{\em JHEP} {\bfseries 08}
  (2013) 046}, \href{http://arxiv.org/abs/0911.4324}{{\ttfamily arXiv:0911.4324
  [hep-th]}}.

\bibitem{Gauntlett:2006af}
J.~P. Gauntlett, O.~A.~P. Mac~Conamhna, T.~Mateos, and D.~Waldram,
  ``{Supersymmetric AdS(3) solutions of type IIB supergravity},''
  \href{http://dx.doi.org/10.1103/PhysRevLett.97.171601}{{\em Phys. Rev. Lett.}
  {\bfseries 97} (2006) 171601},
\href{http://arxiv.org/abs/hep-th/0606221}{{\ttfamily arXiv:hep-th/0606221
  [hep-th]}}.

\bibitem{Gauntlett:2006qw}
J.~P. Gauntlett, O.~A. Mac~Conamhna, T.~Mateos, and D.~Waldram, ``{New
  supersymmetric AdS(3) solutions},''
  \href{http://dx.doi.org/10.1103/PhysRevD.74.106007}{{\em Phys.Rev.}
  {\bfseries D74} (2006) 106007},
\href{http://arxiv.org/abs/hep-th/0608055}{{\ttfamily arXiv:hep-th/0608055
  [hep-th]}}.

\bibitem{Gauntlett:2006ns}
J.~P. Gauntlett, N.~Kim, and D.~Waldram, ``{Supersymmetric AdS(3), AdS(2) and
  Bubble Solutions},''
  \href{http://dx.doi.org/10.1088/1126-6708/2007/04/005}{{\em JHEP} {\bfseries
  04} (2007) 005},
\href{http://arxiv.org/abs/hep-th/0612253}{{\ttfamily arXiv:hep-th/0612253
  [hep-th]}}.

\bibitem{Donos:2008ug}
A.~Donos, J.~P. Gauntlett, and N.~Kim, ``{AdS Solutions Through
  Transgression},'' \href{http://dx.doi.org/10.1088/1126-6708/2008/09/021}{{\em
  JHEP} {\bfseries 0809} (2008) 021},
\href{http://arxiv.org/abs/0807.4375}{{\ttfamily arXiv:0807.4375 [hep-th]}}.

\bibitem{Couzens:2019mkh}
C.~Couzens, H.~het Lam, and K.~Mayer, ``{Twisted $ \mathcal{N} $ = 1 SCFTs and
  their AdS$_{3}$ duals},''
  \href{http://dx.doi.org/10.1007/JHEP03(2020)032}{{\em JHEP} {\bfseries 03}
  (2020) 032}, \href{http://arxiv.org/abs/1912.07605}{{\ttfamily
  arXiv:1912.07605 [hep-th]}}.

\bibitem{Witten:1988ze}
E.~Witten, ``{Topological Quantum Field Theory},''
  \href{http://dx.doi.org/10.1007/BF01223371}{{\em Commun. Math. Phys.}
  {\bfseries 117} (1988) 353}.

\bibitem{Maldacena:2000mw}
J.~M. Maldacena and C.~Nunez, ``{Supergravity description of field theories on
  curved manifolds and a no go theorem},''
  \href{http://dx.doi.org/10.1142/S0217751X01003937}{{\em Int.J.Mod.Phys.}
  {\bfseries A16} (2001) 822--855},
\href{http://arxiv.org/abs/hep-th/0007018}{{\ttfamily arXiv:hep-th/0007018
  [hep-th]}}.

\bibitem{Ferrero:2020laf}
P.~Ferrero, J.~P. Gauntlett, J.~M. P\'erez Ipi\~na, D.~Martelli, and J.~Sparks,
  ``{D3-Branes Wrapped on a Spindle},''
  \href{http://dx.doi.org/10.1103/PhysRevLett.126.111601}{{\em Phys. Rev.
  Lett.} {\bfseries 126} no.~11, (2021) 111601},
  \href{http://arxiv.org/abs/2011.10579}{{\ttfamily arXiv:2011.10579
  [hep-th]}}.

\bibitem{Ferrero:2021etw}
P.~Ferrero, J.~P. Gauntlett, and J.~Sparks, ``{Supersymmetric spindles},''
  \href{http://dx.doi.org/10.1007/JHEP01(2022)102}{{\em JHEP} {\bfseries 01}
  (2022) 102}, \href{http://arxiv.org/abs/2112.01543}{{\ttfamily
  arXiv:2112.01543 [hep-th]}}.

\bibitem{Ferrero:2020twa}
P.~Ferrero, J.~P. Gauntlett, J.~M.~P. Ipi\~na, D.~Martelli, and J.~Sparks,
  ``{Accelerating black holes and spinning spindles},''
  \href{http://dx.doi.org/10.1103/PhysRevD.104.046007}{{\em Phys. Rev. D}
  {\bfseries 104} no.~4, (2021) 046007},
  \href{http://arxiv.org/abs/2012.08530}{{\ttfamily arXiv:2012.08530
  [hep-th]}}.

\bibitem{Colombo:2024mts}
E.~Colombo, S.~M. Hosseini, D.~Martelli, A.~Pittelli, and A.~Zaffaroni,
  ``{Microstates of Accelerating and Supersymmetric AdS4 Black Holes from the
  Spindle Index},''
  \href{http://dx.doi.org/10.1103/PhysRevLett.133.031603}{{\em Phys. Rev.
  Lett.} {\bfseries 133} no.~3, (2024) 031603},
  \href{http://arxiv.org/abs/2404.07173}{{\ttfamily arXiv:2404.07173
  [hep-th]}}.

\bibitem{Cassani:2021dwa}
D.~Cassani, J.~P. Gauntlett, D.~Martelli, and J.~Sparks, ``{Thermodynamics of
  accelerating and supersymmetric AdS4 black holes},''
  \href{http://dx.doi.org/10.1103/PhysRevD.104.086005}{{\em Phys. Rev. D}
  {\bfseries 104} no.~8, (2021) 086005},
  \href{http://arxiv.org/abs/2106.05571}{{\ttfamily arXiv:2106.05571
  [hep-th]}}.

\bibitem{Hosseini:2016tor}
S.~M. Hosseini and A.~Zaffaroni, ``{Large $N$ matrix models for 3d ${\cal N}=2$
  theories: twisted index, free energy and black holes},''
  \href{http://dx.doi.org/10.1007/JHEP08(2016)064}{{\em JHEP} {\bfseries 08}
  (2016) 064},
\href{http://arxiv.org/abs/1604.03122}{{\ttfamily arXiv:1604.03122 [hep-th]}}.

\bibitem{Hosseini:2016ume}
S.~M. Hosseini and N.~Mekareeya, ``{Large $N$ topologically twisted index:
  necklace quivers, dualities, and Sasaki-Einstein spaces},''
  \href{http://dx.doi.org/10.1007/JHEP08(2016)089}{{\em JHEP} {\bfseries 08}
  (2016) 089},
\href{http://arxiv.org/abs/1604.03397}{{\ttfamily arXiv:1604.03397 [hep-th]}}.

\bibitem{Azzurli:2017kxo}
F.~Azzurli, N.~Bobev, P.~M. Crichigno, V.~S. Min, and A.~Zaffaroni, ``{A
  universal counting of black hole microstates in AdS$_{4}$},''
  \href{http://dx.doi.org/10.1007/JHEP02(2018)054}{{\em JHEP} {\bfseries 02}
  (2018) 054},
\href{http://arxiv.org/abs/1707.04257}{{\ttfamily arXiv:1707.04257 [hep-th]}}.

\bibitem{Hosseini:2021fge}
S.~M. Hosseini, K.~Hristov, and A.~Zaffaroni, ``{Rotating multi-charge spindles
  and their microstates},''
  \href{http://dx.doi.org/10.1007/JHEP07(2021)182}{{\em JHEP} {\bfseries 07}
  (2021) 182}, \href{http://arxiv.org/abs/2104.11249}{{\ttfamily
  arXiv:2104.11249 [hep-th]}}.

\bibitem{Boido:2021szx}
A.~Boido, J.~M.~P. Ipi\~na, and J.~Sparks, ``{Twisted D3-brane and M5-brane
  compactifications from multi-charge spindles},''
  \href{http://dx.doi.org/10.1007/JHEP07(2021)222}{{\em JHEP} {\bfseries 07}
  (2021) 222}, \href{http://arxiv.org/abs/2104.13287}{{\ttfamily
  arXiv:2104.13287 [hep-th]}}.

\bibitem{Ferrero:2021ovq}
P.~Ferrero, M.~Inglese, D.~Martelli, and J.~Sparks, ``{Multicharge accelerating
  black holes and spinning spindles},''
  \href{http://dx.doi.org/10.1103/PhysRevD.105.126001}{{\em Phys. Rev. D}
  {\bfseries 105} no.~12, (2022) 126001},
  \href{http://arxiv.org/abs/2109.14625}{{\ttfamily arXiv:2109.14625
  [hep-th]}}.

\bibitem{Couzens:2021rlk}
C.~Couzens, K.~Stemerdink, and D.~van~de Heisteeg, ``{M2-branes on discs and
  multi-charged spindles},''
  \href{http://dx.doi.org/10.1007/JHEP04(2022)107}{{\em JHEP} {\bfseries 04}
  (2022) 107}, \href{http://arxiv.org/abs/2110.00571}{{\ttfamily
  arXiv:2110.00571 [hep-th]}}.

\bibitem{Boido:2022mbe}
A.~Boido, J.~P. Gauntlett, D.~Martelli, and J.~Sparks, ``{Gravitational Blocks,
  Spindles and GK Geometry},''
  \href{http://dx.doi.org/10.1007/s00220-023-04812-8}{{\em Commun. Math. Phys.}
  {\bfseries 403} no.~2, (2023) 917--1003},
  \href{http://arxiv.org/abs/2211.02662}{{\ttfamily arXiv:2211.02662
  [hep-th]}}.

\bibitem{Hosseini:2019iad}
S.~M. Hosseini, K.~Hristov, and A.~Zaffaroni, ``{Gluing gravitational blocks
  for AdS black holes},'' \href{http://dx.doi.org/10.1007/JHEP12(2019)168}{{\em
  JHEP} {\bfseries 12} (2019) 168},
  \href{http://arxiv.org/abs/1909.10550}{{\ttfamily arXiv:1909.10550
  [hep-th]}}.

\bibitem{Faedo:2021nub}
F.~Faedo and D.~Martelli, ``{D4-branes wrapped on a spindle},''
  \href{http://dx.doi.org/10.1007/JHEP02(2022)101}{{\em JHEP} {\bfseries 02}
  (2022) 101}, \href{http://arxiv.org/abs/2111.13660}{{\ttfamily
  arXiv:2111.13660 [hep-th]}}.

\bibitem{Beem:2012mb}
C.~Beem, T.~Dimofte, and S.~Pasquetti, ``{Holomorphic Blocks in Three
  Dimensions},'' \href{http://dx.doi.org/10.1007/JHEP12(2014)177}{{\em JHEP}
  {\bfseries 12} (2014) 177}, \href{http://arxiv.org/abs/1211.1986}{{\ttfamily
  arXiv:1211.1986 [hep-th]}}.

\bibitem{Boido:2022iye}
A.~Boido, J.~P. Gauntlett, D.~Martelli, and J.~Sparks, ``{Entropy Functions For
  Accelerating Black Holes},''
  \href{http://dx.doi.org/10.1103/PhysRevLett.130.091603}{{\em Phys. Rev.
  Lett.} {\bfseries 130} no.~9, (2023) 091603},
  \href{http://arxiv.org/abs/2210.16069}{{\ttfamily arXiv:2210.16069
  [hep-th]}}.

\bibitem{Inglese:2023wky}
M.~Inglese, D.~Martelli, and A.~Pittelli, ``{The spindle index from
  localization},'' \href{http://dx.doi.org/10.1088/1751-8121/ad2225}{{\em J.
  Phys. A} {\bfseries 57} no.~8, (2024) 085401},
  \href{http://arxiv.org/abs/2303.14199}{{\ttfamily arXiv:2303.14199
  [hep-th]}}.

\bibitem{BenettiGenolini:2023kxp}
P.~Benetti~Genolini, J.~P. Gauntlett, and J.~Sparks, ``{Equivariant
  Localization in Supergravity},''
  \href{http://dx.doi.org/10.1103/PhysRevLett.131.121602}{{\em Phys. Rev.
  Lett.} {\bfseries 131} no.~12, (2023) 121602},
  \href{http://arxiv.org/abs/2306.03868}{{\ttfamily arXiv:2306.03868
  [hep-th]}}.

\bibitem{BV:1982}
N.~Berline and M.~Vergne, ``{Classes caract\'{e}ristiques \'{e}quivariantes.
  Formules de localisation en cohomologie \'{e}quivariante},'' {\em C.R. Acad.
  Sc. Paris} {\bfseries 295} (1982) 539--541.

\bibitem{Atiyah:1984px}
M.~F. Atiyah and R.~Bott, ``{The Moment map and equivariant cohomology},''
  \href{http://dx.doi.org/10.1016/0040-9383(84)90021-1}{{\em Topology}
  {\bfseries 23} (1984) 1--28}.

\bibitem{BenettiGenolini:2023yfe}
P.~Benetti~Genolini, J.~P. Gauntlett, and J.~Sparks, ``{Localizing wrapped
  M5-branes and gravitational blocks},''
  \href{http://dx.doi.org/10.1103/PhysRevD.108.L101903}{{\em Phys. Rev. D}
  {\bfseries 108} no.~10, (2023) L101903},
  \href{http://arxiv.org/abs/2308.10933}{{\ttfamily arXiv:2308.10933
  [hep-th]}}.

\bibitem{BenettiGenolini:2023ndb}
P.~Benetti~Genolini, J.~P. Gauntlett, and J.~Sparks, ``{Equivariant
  localization for AdS/CFT},''
  \href{http://dx.doi.org/10.1007/JHEP02(2024)015}{{\em JHEP} {\bfseries 02}
  (2024) 015}, \href{http://arxiv.org/abs/2308.11701}{{\ttfamily
  arXiv:2308.11701 [hep-th]}}.

\bibitem{BenettiGenolini:2024kyy}
P.~Benetti~Genolini, J.~P. Gauntlett, Y.~Jiao, A.~L\"uscher, and J.~Sparks,
  ``{Localization and attraction},''
  \href{http://dx.doi.org/10.1007/JHEP05(2024)152}{{\em JHEP} {\bfseries 05}
  (2024) 152}, \href{http://arxiv.org/abs/2401.10977}{{\ttfamily
  arXiv:2401.10977 [hep-th]}}.

\bibitem{Couzens:2024vbn}
C.~Couzens and A.~L\"uscher, ``{A geometric dual of F-maximization in massive
  type IIA},'' \href{http://dx.doi.org/10.1007/JHEP08(2024)218}{{\em JHEP}
  {\bfseries 08} (2024) 218}, \href{http://arxiv.org/abs/2406.15547}{{\ttfamily
  arXiv:2406.15547 [hep-th]}}.

\bibitem{Hristov:2024cgj}
K.~Hristov, ``{Equivariant localization and gluing rules in 4d $\mathcal{N}=2$
  higher derivative supergravity},''
  \href{http://arxiv.org/abs/2406.18648}{{\ttfamily arXiv:2406.18648
  [hep-th]}}.

\bibitem{BenettiGenolini:2024xeo}
P.~Benetti~Genolini, J.~P. Gauntlett, Y.~Jiao, A.~L\"uscher, and J.~Sparks,
  ``{Localization of the Free Energy in Supergravity},''
  \href{http://dx.doi.org/10.1103/PhysRevLett.133.141601}{{\em Phys. Rev.
  Lett.} {\bfseries 133} no.~14, (2024) 141601},
  \href{http://arxiv.org/abs/2407.02554}{{\ttfamily arXiv:2407.02554
  [hep-th]}}.

\bibitem{Cassani:2024kjn}
D.~Cassani, A.~Ruip\'erez, and E.~Turetta, ``{Localization of the 5D
  supergravity action and Euclidean saddles for the black hole index},''
  \href{http://dx.doi.org/10.1007/JHEP12(2024)086}{{\em JHEP} {\bfseries 12}
  (2024) 086}, \href{http://arxiv.org/abs/2409.01332}{{\ttfamily
  arXiv:2409.01332 [hep-th]}}.

\bibitem{BenettiGenolini:2024hyd}
P.~Benetti~Genolini, J.~P. Gauntlett, Y.~Jiao, A.~L\"uscher, and J.~Sparks,
  ``{Toric gravitational instantons in gauged supergravity},''
  \href{http://dx.doi.org/10.1103/PhysRevD.111.046024}{{\em Phys. Rev. D}
  {\bfseries 111} no.~4, (2025) 046024},
  \href{http://arxiv.org/abs/2410.19036}{{\ttfamily arXiv:2410.19036
  [hep-th]}}.

\bibitem{Crisafio:2024fyc}
M.~K. Crisafio, A.~Fontanarossa, and D.~Martelli, ``{Nuts, bolts and
  spindles},'' \href{http://dx.doi.org/10.1007/s11005-025-01915-2}{{\em Lett.
  Math. Phys.} {\bfseries 115} no.~2, (2025) 27},
  \href{http://arxiv.org/abs/2412.00428}{{\ttfamily arXiv:2412.00428
  [hep-th]}}.

\bibitem{BenettiGenolini:2024lbj}
P.~Benetti~Genolini, J.~P. Gauntlett, Y.~Jiao, A.~L\"uscher, and J.~Sparks,
  ``{Equivariant localization for $D=4$ gauged supergravity},''
  \href{http://arxiv.org/abs/2412.07828}{{\ttfamily arXiv:2412.07828
  [hep-th]}}.

\bibitem{Hristov:2021qsw}
K.~Hristov, ``{4d $ \mathcal{N} $ = 2 supergravity observables from
  Nekrasov-like partition functions},''
  \href{http://dx.doi.org/10.1007/JHEP02(2022)079}{{\em JHEP} {\bfseries 02}
  (2022) 079}, \href{http://arxiv.org/abs/2111.06903}{{\ttfamily
  arXiv:2111.06903 [hep-th]}}.

\bibitem{Martelli:2023oqk}
D.~Martelli and A.~Zaffaroni, ``{Equivariant localization and holography},''
  \href{http://dx.doi.org/10.1007/s11005-023-01752-1}{{\em Lett. Math. Phys.}
  {\bfseries 114} no.~1, (2024) 15},
  \href{http://arxiv.org/abs/2306.03891}{{\ttfamily arXiv:2306.03891
  [hep-th]}}.

\bibitem{Colombo:2023fhu}
E.~Colombo, F.~Faedo, D.~Martelli, and A.~Zaffaroni, ``{Equivariant volume
  extremization and holography},''
  \href{http://dx.doi.org/10.1007/JHEP01(2024)095}{{\em JHEP} {\bfseries 01}
  (2024) 095}, \href{http://arxiv.org/abs/2309.04425}{{\ttfamily
  arXiv:2309.04425 [hep-th]}}.

\bibitem{Colombo:2025ihp}
E.~Colombo, V.~Dimitrov, D.~Martelli, and A.~Zaffaroni, ``{Equivariant
  localization in supergravity in odd dimensions},''
  \href{http://arxiv.org/abs/2502.15624}{{\ttfamily arXiv:2502.15624
  [hep-th]}}.

\end{thebibliography}

\providecommand{\href}[2]{#2}\begingroup\raggedright\endgroup

\end{document}